\documentclass[conference]{IEEEtran}

\usepackage{caption}
\usepackage{subcaption}
\usepackage{graphicx}
\usepackage{tikz}
\usepackage{amsmath, amsfonts, amsthm, amssymb}
\usepackage{xcolor}
\usepackage[ruled, vlined, linesnumbered]{algorithm2e}
\usepackage{pifont}
\usepackage{makecell,rotating}
\usepackage{algpseudocode}
\usepackage{upgreek}
\usepackage{eucal}
\usepackage{tcolorbox}
\tcbuselibrary{breakable, skins}
\usepackage{threeparttable}
\usepackage{lipsum}
\usepackage{verbatim}
\usepackage{setspace}
\usepackage{multirow}
\usepackage{booktabs}
\usepackage{tablefootnote}
\usepackage{tabularx}
\usepackage{bbding,enumitem}
\usepackage{balance}
\usepackage{array}
\usepackage{url}
\usepackage{cite}
\usepackage[hidelinks]{hyperref}
\usepackage{tcolorbox}

\newcommand{\ie}{{i.e.}}
\newcommand{\eg}{{e.g.}}

\newcommand{\ours}{\textsc{AgentPrint}}
\newcommand{\gpts}{\textit{GPTs}}
\begin{document}

\title{
    Exposing LLM User Privacy via Traffic Fingerprint Analysis: A Study of Privacy Risks in LLM Agent Interactions
}
\author{
    \IEEEauthorblockN{
        Yixiang Zhang\IEEEauthorrefmark{1},
        Xinhao Deng\IEEEauthorrefmark{2}\IEEEauthorrefmark{1},
        Zhongyi Gu\IEEEauthorrefmark{1},
        Yihao Chen\IEEEauthorrefmark{1},
        Ke Xu\IEEEauthorrefmark{1}
        Qi Li\IEEEauthorrefmark{1},
        Jianping Wu\IEEEauthorrefmark{1}
    }
    \IEEEauthorblockA{
        \IEEEauthorrefmark{1}Tsinghua University
        \IEEEauthorrefmark{2}Ant Group
    }
    \IEEEauthorblockA{
        \{zhangyix24, gu-zy24, yh-chen21\}@mails.tsinghua.edu.cn
    }
    \IEEEauthorblockA{
        \{dengxinhao, xuke, qli01\}@tsinghua.edu.cn,
        jianping@cernet.edu.cn
    }
}

\IEEEoverridecommandlockouts
\makeatletter\def\@IEEEpubidpullup{6.5\baselineskip}\makeatother

\maketitle

\begin{abstract}
Large Language Models (LLMs) are increasingly deployed as agents that orchestrate tasks and integrate external tools to execute complex workflows. We demonstrate that these interactive behaviors leave distinctive fingerprints in encrypted traffic exchanged between users and LLM agents. By analyzing traffic patterns associated with agent workflows and tool invocations, adversaries can infer agent activities, distinguish specific agents, and even profile sensitive user attributes. To highlight this risk, we develop \ours{}, which achieves an F1-score of 0.866 in agent identification and attains 73.9\% and 69.1\% top-3 accuracy in user attribute inference for simulated- and real-user settings, respectively. These results uncover an overlooked risk: the very interactivity that empowers LLM agents also exposes user privacy, underscoring the urgent need for technical countermeasures alongside regulatory and policy safeguards.
\end{abstract}

\section{Introduction}
In recent years, the concept of AI agents has gained increasing prominence, encompassing systems that range from task-specific applications to general-purpose assistants.
Within this broad spectrum, AI agents powered by Large Language Models (LLMs), also referred to as LLM agents, have emerged as a representative and influential class~\cite{survey_agents, survey_agent_rise}. 
Equipped with integrated perception and execution components, these agents have evolved from simple prompt responders (\ie, chatbots) into autonomous assistants capable of planning, remembering, and interacting across digital systems, enabling the handling of complex, multi-step tasks across various domains~\cite{survey_agents, survey_agent_rise,survey_llm, survey_llm_tool1, survey_llm_tool2, survey_agent_evaluation_1, survey_agent_evaluation_2, survey_agent_deep_research}. 
As a result, users are increasingly able to delegate labor-intensive and time-consuming workflows to LLM agents tailored for specialized tasks, such as research~\cite{Agent_for_study}, software development~\cite{Agent_software}, customer support~\cite{LLM_custom_support}, disease diagnosis~\cite{Agent_medical}, and personal productivity~\cite{LLM_personal_Productivity}. 

However, the rapid advancement and increasing deployment of LLM agents have simultaneously raised serious concerns regarding the security risks inherent in this emerging paradigm. Built on the language understanding and generation capabilities of LLMs, LLM agents inherit and even amplify a range of content-related risks. These include adversarial prompts~\cite{LLM_Jailbreak,LLM_hijacking,LLM_Jailbreak2,LLM_Jailbreak3}, hallucinations~\cite{LLM_hallucinations,LLM_hallucinations2}, and harmful outputs such as bias and toxicity~\cite{LLM_bias,LLM_bias2}. Recent research has primarily examined how these risks manifest in the agent setting, particularly as agents engage with external environments through content-driven perception and execution modules~\cite{survey_agents}. For instance, malicious inputs can be exploited to mislead agents~\cite{sec_gpts_tracker, sec_agent_jailbreak}, sensitive information may be exfiltrated through unauthorized tool invocations~\cite{sec_agent_privacy}, and agent workflows can be manipulated by adversarial environmental feedback~\cite{sec_agent_dynamic_env_attack, sec_agent_dynamic_env_attack2}.

These threats primarily arise from unsafe content originating from diverse sources, including malicious user inputs, unauthorized tool payloads, and adversarial environmental signals. Yet beyond these well-studied content-driven threats, a critical but often overlooked risk emerges from the unique capabilities of LLM agents: novel network-level side-channel~\cite{sec_side_channel} vulnerabilities. Specifically, user privacy can be inferred from encrypted end-to-end traffic generated during user-agent interactions. The proliferation of advanced, task-specific LLM agents thus undermines the confidentiality guarantees traditionally offered by encryption in user-agent communications.

In this article, we demonstrate that task-specific LLM agents leak information through network side-channels by leaving unique fingerprints (\ie, traffic patterns) of network traffic. These fingerprints create a new and overlooked privacy attack surface. As shown in Figure~\ref{fig:main-1-overview}A, by analyzing fingerprints generated during interactions between a user and an LLM agent, an adversary can reliably infer the behaviors the agent is exhibiting, identify the specific agent in use, and even reconstruct sensitive information about the user, such as the occupational role at the category level. For instance, a doctor’s occupation could be inferred by identifying the use of a disease diagnosis LLM agent through network traffic analysis. In other words, while existing encryption techniques protect the content of messages, they do not safeguard the interaction patterns that define LLM agents. This finding challenges the trust in encryption in current privacy protections and introduces a new class of risks unique to the era of LLM agents.

Our study delivers three key insights and technical contributions. First, we show that user-agent interactions inherently generate distinctive and measurable traffic fingerprints. Second, we empirically demonstrate that these fingerprints can be leveraged to classify agent behaviors, identify specific agents, and infer user attributes. Third, we highlight a critical implication: the very operational characteristics that underpin the utility of LLM agents simultaneously introduce novel privacy risks for users, \eg, user attribute inference achieves a top-3 accuracy of 73.9\%. Collectively, these findings emphasize the urgent need to rethink both technical safeguards and regulatory frameworks as LLM agents become deeply embedded in everyday life.

\section{Observing Traffic Patterns during Agent Interaction}
\begin{figure*}
  \centering
  \includegraphics[]{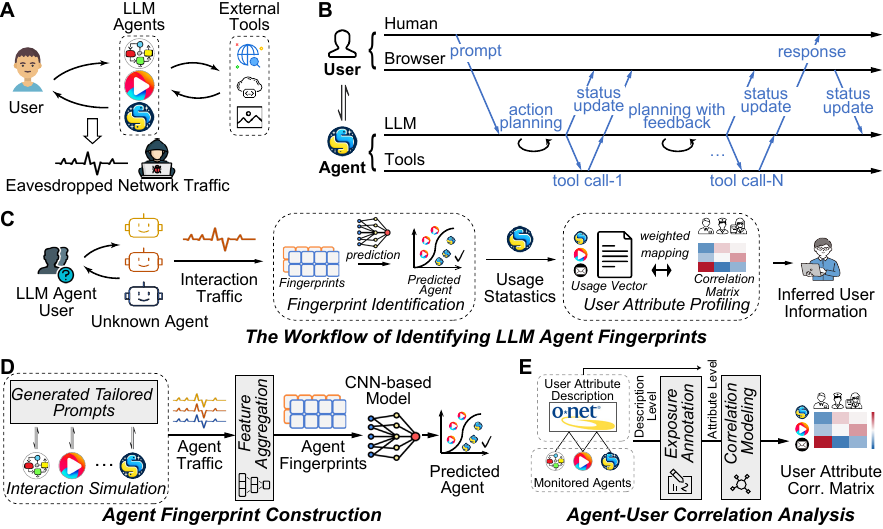}
  \caption{
  \textbf{Overview of \ours.} 
  (\textbf{A}) \textbf{Threat model}: an adversary can uncover private user information by eavesdropping and analyzing traffic generated during interactions with LLM-based AI agents.
  (\textbf{B}) \textbf{Unpacking User-LLM Agent Interaction}: an LLM agent autonomously plans and performs tool calls based on user prompts, integrating feedback to adapt its behavior and generating a response to the user. The browser and the LLM vendor synchronize and update states during this process. The behaviors of the LLM agent result in distinctive network traffic patterns.
  (\textbf{C}) \textbf{The workflow of identifying LLM agent fingerprints}: the adversary initially extracts and identifies traffic fingerprints to infer agent-level behaviors, and subsequently infers sensitive user-level information by applying an agent-user attribute correlation matrix to aggregated agent usage over time.
  (\textbf{D}) \textbf{Agent Fingerprint Construction}: the adversary designs tailored functional prompts and simulates user-agent interactions to generate diverse, fingerprint-enriched traffic. It then extracts and aggregates traffic features to construct agent fingerprints and trains a CNN-based classifier to identify the agent interacting with the user.
  (\textbf{E}) \textbf{Agent-User Correlation Analysis}: the adversary gathers sensitive user-attribute descriptions and agent information, annotates agent exposures with respect to these attributes, and then models agent-attribute correlations to construct an agent-user attribute correlation matrix.
  }
  \label{fig:main-1-overview}
\end{figure*}

We consider an adversary capable of passively monitoring encrypted network traffic between a user and LLM agents~(Figure~\ref{fig:main-1-overview}A). Although encryption protects the message contents, metadata such as packet sizes and transmission timing remain visible. For conventional text-based chatbots, this leakage reveals little beyond conversation length or frequency. In contrast, user interactions with LLM agents exhibit fundamentally more characteristics. These agents handle diverse inputs, maintain multi-step workflows, and orchestrate external tools to accomplish tailored tasks~(Figure~\ref{fig:main-1-overview}B). Such interactions are characterized by two defining properties: \textit{multimodality} and \textit{processuality}. 
 
\textit{Multimodality} arises from the agent’s ability to process and generate diverse input-output forms, including images, code, generated files, and structured responses from third-party APIs. The heterogeneity of these outputs, both in size and composition, produces traffic bursts and packet sequences that are strongly dependent on the underlying modality.

\textit{Processuality} captures the sequential workflows that emerge from tool use. It manifests in two dimensions: (1) the organization of stepwise actions, and (2) the execution characteristics of individual actions, such as API calls with distinct time delays or multi-stage reasoning.
Each dimension is characterized by specific latencies and response patterns. These dynamics leave recognizable signatures in traffic volume and timing.

The properties above mark a clear departure from conventional LLM services: while encryption protects textual content, it cannot obscure the fingerprints introduced by distinctive interaction patterns of LLM agents. As illustrated in Figure~\ref{fig:main-1-overview}, these fingerprints can be leveraged throughout the attack lifecycle, activated during training, monitored during attack, attributed to specific agents, and ultimately correlated with sensitive user attributes.

\begin{figure*}
	\centering
	\includegraphics[width=\textwidth]{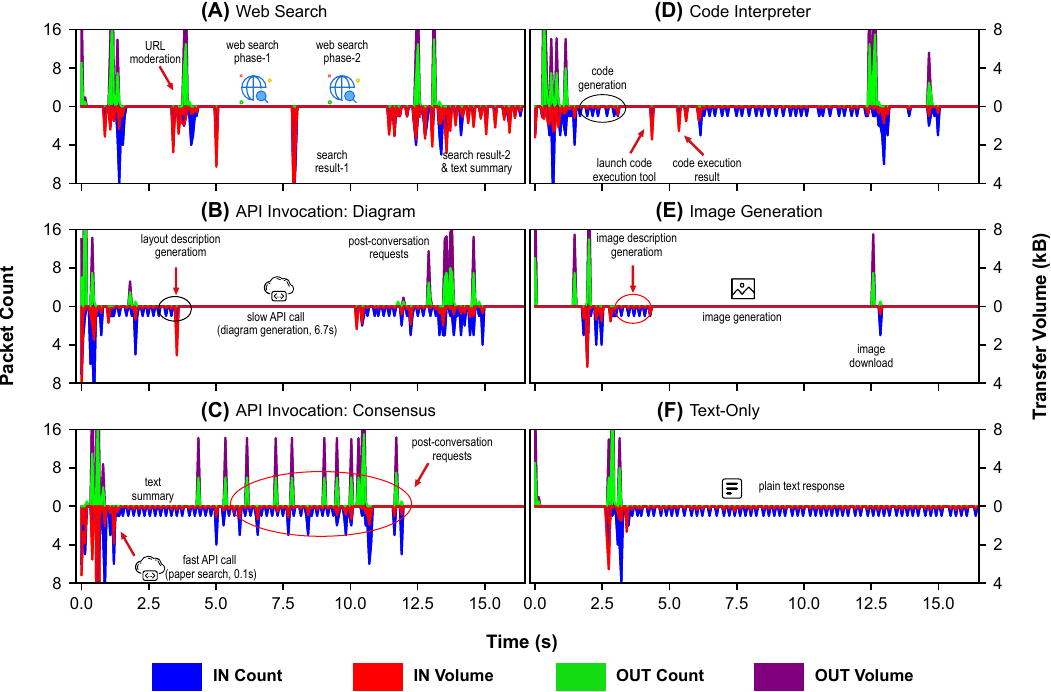}
	\caption{\textbf{Traffic signatures of typical LLM agent behaviors, illustrated through six representative examples.} 
    (\textbf{A}) Two-phase web search for retrieving and summarizing papers on \texttt{science.org}. 
    (\textbf{B}) Diagram creation via a slow third-party API call. 
    (\textbf{C}) Literature search via a rapid third-party API call. 
    (\textbf{D}) Python code generation followed by remote execution. 
    (\textbf{E}) Image generation followed by asset download. 
    (\textbf{F}) Text-only dialogue without tool use. 
    Each figure shows four curves corresponding to multi-view traffic features, which together reveal distinctive interaction patterns that can result in traffic fingerprints.
    }
	\label{fig:main-2-observation}
\end{figure*}

To assess whether such fingerprints arise in practice, we collect traffic traces spanning diverse LLM agent behaviors. Figure~\ref{fig:main-2-observation} presents six representative cases. Distinctive patterns emerge clearly: a two-phase web search exhibits bursts aligned with retrieval and subsequent summarization~(Figure~\ref{fig:main-2-observation}A); diagram creation first produces a textual layout description, followed by a single API call encapsulating the full layout, then a prolonged idle interval and several client-side post-conversation requests~(B); literature retrieval triggers a fast initial API call immediately followed by sustained textual summarization, again concluding with multiple post-conversation requests~(C); code generation with remote execution involves a prolonged multi-step exchange~(D); image generation begins with a textual description, introduces a waiting period, and culminates in a large outbound burst~(E); and text-only dialogue yields relatively uniform traffic flows~(F).

Notably, these distinctions are observed without access to the encrypted payload. Timing and volume alone are sufficient to differentiate behaviors. Each figure presents four complementary feature views capturing packet counts and transmission volumes at multiple granularities, which together expose stable and characteristic signatures of agent activity.

A broader statistical analysis across additional agents confirms the widespread presence of these fingerprints~(see Supplementary Materials). In conclusion, these findings demonstrate that agent interactions inherently leak recognizable traffic signatures, thereby motivating our subsequent investigation into how such fingerprints may be systematically exploited.

\section{Collecting Fingerprints of LLM Agents}
\begin{figure*}
  \centering
  \includegraphics[]{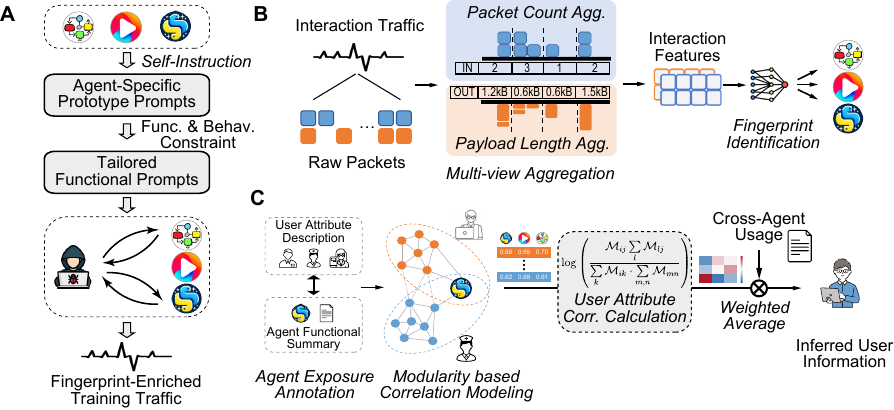}
  \caption{
    \textbf{Analytical framework for fingerprinting LLM agents from interaction traffic and downstream occupation profiling.} 
    (\textbf{A}) Prototype prompts are produced by self-instruction and constrained to elicit representative agent behaviors; they are then used to simulate user-agent interactions that form fingerprint-enriched training traffic.
    (\textbf{B}) Interaction traffic is initially extracted as packet sequences, and then aggregated by packet count and payload length within fixed-gap time windows, producing distinctive fingerprints that enable CNN-based recognition of agent behaviors and identities.
    (\textbf{C}) Monitored agents are annotated by assessing the alignment of their functionalities with detailed user-attribute descriptions, supporting network modularity-based agent-attribute correlation modeling. Comparative-advantage correlation scores are then derived and combined with cross-agent usage to infer user information.
    This framework illustrates how agent-specific interaction patterns can be systematically exposed and exploited for fingerprinting and downstream user attribute inference.
    }
  \label{fig:main-3-design}
\end{figure*}

Conceptually, an adversary could monitor user-agent traffic, extract distinct fingerprints, and associate them with specific agents or downstream user attributes~(Figure~\ref{fig:main-1-overview}C). To assess this possibility, we develop an experimental workflow that captures representative agent interactions to generate training data, encodes their traffic dynamics, and tests whether the resulting fingerprints enable reliable inferences.

A key prerequisite for this workflow is to collect traffic traces that consistently reveal each agent's characteristic behaviors. The challenge is that agents do not always exhibit such behaviors under arbitrary queries, leading to generic exchanges with limited fingerprints. For instance, an email-writing agent might respond to a casual health question using only its base language generating capabilities, but such an interaction neither engages its core functionalities nor produces distinctive traffic patterns. Therefore, it is essential to design interactions that can reliably activate these core functions for fingerprint analysis. This requirement carries two important implications. First, it ensures that the classifier learns informative features that distinguish agent-specific traffic. Second, it aligns the identification results with the agent's core functionalities, providing downstream profiling analysis with both practicality and interpretability.

To ensure that traffic traces consistently reveal agent-specific fingerprints, we develop a two-stage prompt construction strategy~(Figure~\ref{fig:main-3-design}A). In the introspection stage, agents are prompted to generate prototype queries reflecting their declared capabilities (\eg, summarizing video content, generating websites, or retrieving literature). In the observation stage, these prototypes are refined based on metadata and interaction patterns derived from agent documentation and case studies, incorporating functional and behavioral constraints. The resulting tailored prompts reliably trigger each agent's core functions and elicited realistic responses, producing traffic traces matching their intended use. This approach enhances both the diversity and authenticity of the collected traffic, in contrast to naïve prompting, which often leads to generic exchanges. Further methodological details are provided in the Supplementary Materials.

Building on these curated interactions, we construct a multi-view traffic feature representation~(Figure~\ref{fig:main-3-design}B). Each traffic flow is segmented into fixed-length time slices, from which two complementary metrics are extracted per slice: packet counts, capturing the structural and temporal dynamics driven by process workflows, and transmission volumes, reflecting the variability introduced by multimodal outputs. By combining these features across multiple granularities, the feature matrix encodes the core properties, \ie, processuality and multimodality, which underlie the fingerprints identified in Figure~\ref{fig:main-2-observation}. Crucially, this representation relies solely on metadata observable from encrypted traffic, making it directly applicable in real-world adversarial settings. Implementation details can be found in the Supplementary Materials.

\begin{figure*}
	\centering
	\includegraphics[]{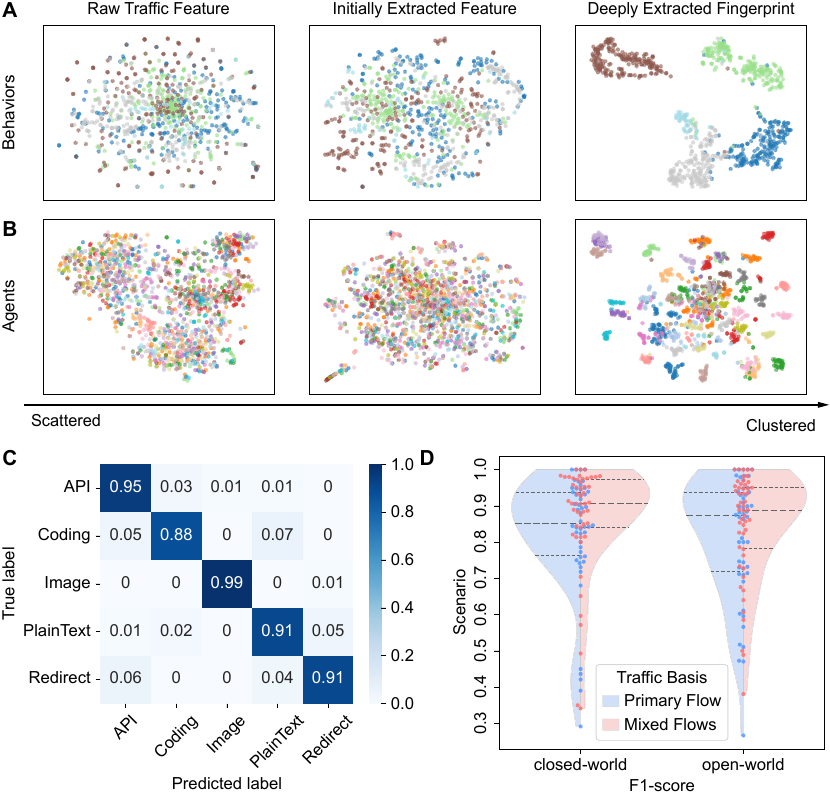}
	\caption{\textbf{Performance of LLM agent behavior and identity classification from user-agent interaction traffic.} 
    (\textbf{A}) t-SNE visualization of agent behavior recognition across model-learned representations, spanning raw traffic to progressively extracted deep fingerprints.
    (\textbf{B}) t-SNE visualization of agent identity recognition across model-learned representations.
    Both visualizations demonstrate the evolution from scattered distributions to well-clustered groups.
    (\textbf{C}) Confusion matrix of behavior classification across five categories. 
    (\textbf{D}) Violin plots of F1-score distributions for identity recognition of 50 monitored agents under four settings, combining closed- vs. open-world scenarios of available agents with primary- vs. mixed-flow traffic.
    These results confirm that both agent behaviors and identities can be reliably inferred from encrypted traffic, underscoring the feasibility of traffic-based fingerprinting.
    }
	\label{fig:main-4-traffic_result}
\end{figure*}

We now assess whether these fingerprints are sufficient to differentiate agent behaviors and individual agent identities. As an empirical test, we train a supervised CNN-based classifier~\cite{wf_rf} on the extracted features and evaluate the separability of these features across behavior and identity categories~(Figure~\ref{fig:main-4-traffic_result}), which aims to determine whether the traffic traces themselves are highly informative for classification. These results confirm that agent fingerprints can be reliably utilized for inference. A t-SNE projection of the traffic features reveals well-defined clusters for both the five agent behaviors~(Figure~\ref{fig:main-4-traffic_result}A) and 50 monitored LLM agents~(Figure~\ref{fig:main-4-traffic_result}B). Direct classification achieves over 90\% accuracy for most behaviors, with the confusion matrix showing minimal cross-category errors~(Figure~\ref{fig:main-4-traffic_result}C). Identity inference is also reliable, \eg, for the 50 monitored agents, median F1-scores exceed 0.85 under closed-world assumptions and stay above 0.80 in open-world scenarios involving previously unseen agents. Incorporating mixed traffic flows, such as third-party resource loading, further enhances accuracy, highlighting the stability of fingerprints across different interaction contexts. Violin plots of F1-score distributions~(Figure~\ref{fig:main-4-traffic_result}D) demonstrate consistent performance across all four settings, combining closed- vs. open-world scenarios with primary- vs. mixed-flow traffic (see Supplementary Materials). Overall, the classification experiment yields 0.924 macro F1-score and 94.1\% accuracy for agent behavior identification, and 0.866 macro F1-score and 86.7\% accuracy for agent identity recognition in the closed-world scenario with mixed-flow traffic. 

Therefore, we conclude that the distinctive interaction fingerprints observed in Figure~\ref{fig:main-2-observation} are not only visible but also exploitable. Even without payload access, traffic patterns suffice to reveal both the type of behaviors being executed and the specific agent in use, which holds consistently across multiple experimental settings and ablation analysis.

\section{Revealing LLM Risks at the User Level}

The fingerprints revealed by agent interactions go beyond identifying behaviors or individual agents. Over time, they accumulate into stable usage patterns that can expose personal information, including demographics, socioeconomic status (SES), and sensitive data such as health conditions or political orientation (table~\ref{tab:privacy_risk}). For simplicity, we focus on occupation as a representative case study, because occupation not only is a key component of SES~\cite{privacy_occ1, privacy_occ2, FTC2024}, but also benefits from comprehensive datasets and established taxonomies~\cite{onet2023}, which allow for a rigorous evaluation~\cite{llm_occ2023, llm_occ2024, llm_occ_ms2024} and makes occupational profiling both societally significant and analytically manageable.

We leverage cross-agent usage (\ie, statistics of used agents over time) as a signal for occupational profiling. Different professions tend to rely on distinct groups of agents. For example, software engineers are likely to frequently use coding and debugging agents, while medical researchers are more prone to engage with literature search and medical knowledge base agents. 
To capture these profession-specific patterns, we embed agents into an occupation network, where occupations are linked by shared Detailed Work Activities and clustered as communities~\cite{GPT_labor_science}. By integrating the observed agent functionalities with this network representation, we construct a correlation matrix that maps agent usage patterns to occupational categories~(Figure~\ref{fig:main-3-design}C). Methodological details are provided in the Supplementary Materials.

\begin{figure*}
	\centering
	\includegraphics[]{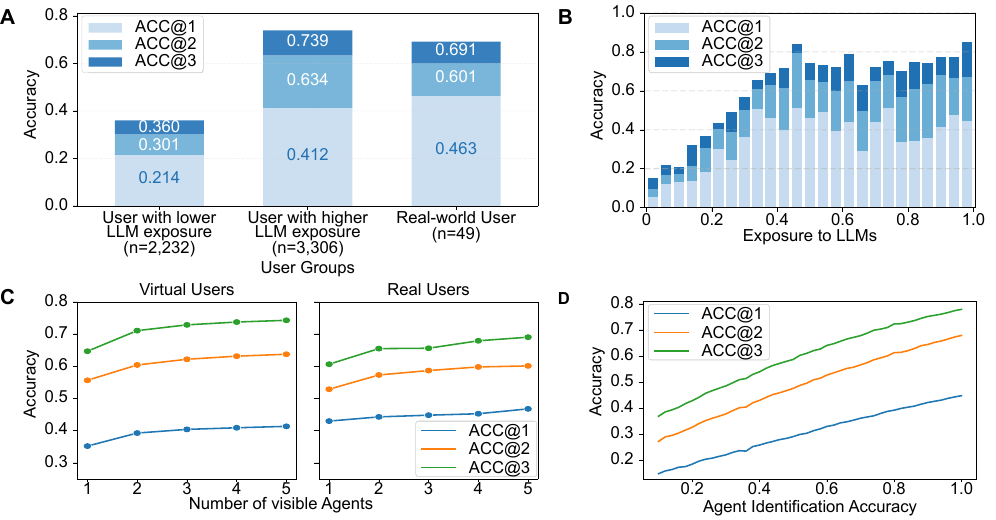}
	\caption{\textbf{User occupation profiling by LLM agent usage.}  
    (\textbf{A}) Top-$K$ ($K$=1, 2, 3) accuracy for virtual user with lower ($<$ 0.4) and higher ($\geq$ 0.4) LLM exposure~\cite{GPT_labor_science}, and real users.
    (\textbf{B}) Effect of occupational LLM exposure level on inference accuracy for virtual users. Accuracy increases steadily with exposure up to 0.4 and then stabilizes around 0.7, indicating that highly exposed occupations are more vulnerable to profiling. 
    (\textbf{C}) Effect of the number of visible agents on inference accuracy. Accuracy improves as more agents are observed, for both virtual and real users. 
    (\textbf{D}) Dependence of occupation profiling on agent identification accuracy. Inference accuracy grows almost linearly with upstream identification accuracy, showing that reliable agent recognition directly enhances downstream profiling. 
    Together, these results demonstrate the feasibility of inferring user occupation categories from agent usage patterns and highlight the associated privacy risks.
    }
	\label{fig:main-5-occupation_result}
\end{figure*}

We assess the effectiveness of this approach using both simulated and real users. Simulated users are generated from the O*NET dataset~\cite{onet2023}, which defines 923 occupations and associated tasks, with interaction traces collected from 55 representative agents. Occupations are grouped into 12 category-level communities, and the goal is to infer each user's category based on traffic during agent use. As shown in Figure~\ref{fig:main-5-occupation_result}A, for 3{,}306 users from 551 occupations with higher exposure to LLMs, occupational inference achieves a top-3 accuracy of 73.9\%. When extended to all 923 occupations~(N=5{,}538), the accuracy drops to 58.9\%, as discriminability decreases for roles with lower exposure. To validate these results in a real-world context, we conduct a study with 49 participants who had prior experience using LLM agents. Their usage preferences yield a comparable top-3 accuracy of 69.1\%. Full numerical results are provided in Supplementary tables~\ref{tab:infer_virtual_user} and \ref{tab:infer_real_user}.

Figure~\ref{fig:main-5-occupation_result} also illustrates the mechanisms driving these results. For simulated users, accuracy increases steadily with the degree of occupational exposure to LLMs, then levels off once exposure surpasses a certain threshold (see Figure~\ref{fig:main-5-occupation_result}B). Accuracy also improves as more agents appear in a user's interaction history, a trend observed in both simulated and real users, highlighting the cumulative value of cross-agent evidence~(Figure~\ref{fig:main-5-occupation_result}C). Finally, inference accuracy shows a near-linear increase with the accuracy of upstream agent identification, indicating that reliable agent fingerprinting directly enhances downstream occupational profiling~(Figure~\ref{fig:main-5-occupation_result}D). Additional analyses are provided in the Supplementary Materials.

Together, these findings show that traffic fingerprints not only compromise individual sessions but can be aggregated to reveal persistent personal attributes. Occupational profiling exemplifies a broader principle, \ie, traffic patterns, even with encrypted message contents, are sufficient to expose sensitive aspects of user identity. What begins as fingerprints of agent behavior thus evolves into long-term privacy risks for users in the era of LLM agents.

\section{Rethinking Privacy Leaks Induced by LLM Agents}
The emergence of LLM-based agents marks a fundamental shift from traditional chatbot services. Rather than acting as passive text responders, these agents operate as autonomous systems capable of planning tasks, invoking tools, and interacting with the physical and digital world. Our study shows that these very capabilities, which enhance agents' power and versatility, also leave behind distinctive traffic fingerprints. Multi-stage workflows, delays characteristic of tool calls, and multimodal payloads of varying sizes collectively create observable patterns even within encrypted traffic. These fingerprints are sufficiently consistent to allow inference of behaviors, agent identities, and, ultimately, long-term user attributes. In this way, the evolution from static dialogue systems to interactive agents has unintentionally introduced a new privacy attack surface.

The implications extend far beyond a technical demonstration. In sensitive domains such as healthcare~\cite{llm_medicine1, llm_medicine2, llm_medicine3, llm_medicine4, llm_medicine5, llm_medicine6}, education~\cite{llm_education1, llm_education2, llm_education3}, and government~\cite{llm_government1, llm_government2}, which are increasingly integrated with LLMs and agent services, fingerprinting could be exploited to expose or discriminate against individuals based on the nature of their agent interactions. For example, identifying a user’s repeated engagement with diagnostic agents could reveal medical conditions; monitoring research-related agents might expose intellectual property or institutional priorities; and tracking student interactions could be misused to restrict access to educational resources. In the private sector, such information could fuel competitive intelligence~\cite{privacy_book, FTC2024}, with companies monitoring competitors' use of specialized agents to infer ongoing projects or business strategies. These scenarios highlight that while agent fingerprints may seem subtle, their misuse can lead to significant privacy risks and far-reaching societal consequences.

Addressing these risks requires both technical and regulatory responses. On the technical side, mitigation strategies include injecting dummy packets~\cite{wf_front, wf_wtfpad}, batching transmissions~\cite{sec_llm_prompt}, or shaping traffic to obscure behavioral signatures~\cite{wf_tamaraw, wf_walkie}. Each approach involves trade-offs for overhead, latency, and usability, and further research is needed to find an optimal balance between protection and performance. We present an initial analysis of these strategies in the Supplementary Materials, but a comprehensive evaluation remains an open challenge. From a regulatory perspective, service providers should be mandated to disclose traffic-based privacy risks in their terms of service, while policymakers should update privacy frameworks to explicitly account for side-channel leakage in agent-driven systems. Without such measures, users will remain unaware of risks that encryption alone cannot mitigate.

Finally, our study has limitations that suggest several avenues for future research. We analyze a representative, yet finite, set of agents, leaving open the question of how fingerprints generalize across the broader ecosystem of community-developed agents. While the presence of more agents may increase classification complexity, it could also introduce richer signals of user attributes, potentially enhancing profiling capabilities. Meanwhile, there is a growing trend of GUI agents~\cite{openai_chatgpt_agent, sec_gui_agent} that delegate user-local applications to LLMs via video streams. However, this paradigm still poses the risk that user behaviors may be inferred from the patterns of uploading video streams~\cite{wf_video} and from the action sequences returned by the agents. Although we focus on occupation as a key user attribute, other dimensions, such as personal interests, political orientations, or health status, may also be vulnerable.
Our real-world user study is constrained in scale, not only by budget and time limitations, but also by the inherent challenges of prolonged data collection and intensive processing.
Additionally, our experiments assume an adversary capable of passively monitoring traffic at the network level; future research should explore more sophisticated adversaries with active manipulation capabilities, as well as those with more constrained access. These open questions highlight the need for broader collaboration across the security, AI, and policy communities.

These considerations above highlight a broader lesson. The operational characteristics that make LLM agents powerful also create new privacy vulnerabilities. As agents become increasingly integrated into everyday workflows, addressing these risks will require not only technical innovation but also institutional vigilance and regulatory foresight.

\section{Concluding \ours}

We uncover a critical yet often overlooked aspect of user privacy in emerging LLM agent services. Unlike text-only chatbots, agents function through interactive workflows and multimodal tool usage, leaving distinctive fingerprints in the encrypted traffic exchanged between users and agents. These fingerprints can be exploited to infer agent behaviors, identify specific agents, and ultimately reconstruct sensitive user attributes, such as occupational roles.

Our findings challenge the prevailing assumption that encryption alone ensures privacy, demonstrating instead that the very capabilities that empower LLM agents also introduce new vulnerabilities for users. Addressing these risks will require both technical safeguards to mitigate traffic-based leakage and regulatory frameworks to ensure transparency and accountability. As LLM agents become increasingly integrated into everyday workflows, rethinking privacy in this new paradigm is not just optional but essential.


\bibliographystyle{IEEEtran}
\bibliography{references}

\appendices
\section{Literature Review} 
\subsection{Large Language Models}
In recent years, large language models (LLMs) have become central to advances in Artificial Intelligence (AI) and Natural Language Processing (NLP)~\cite{survey_llm}. These models, comprising hundreds of millions to hundreds of billions of parameters, are capable of both understanding and generating human language. At their core, LLMs operate by predicting the next token in a sequence given its preceding context, a mechanism that underpins the majority of natural language generation tasks~\cite{survey_llm}.

Most LLMs are built on the Transformer architecture~\cite{transformer}, which uses self-attention mechanisms and stacked encoder-decoder blocks to capture complex semantic relationships within token sequences. Following tokenization and embedding, inputs are processed through multiple transformer layers, producing a probability distribution over the vocabulary. The tokens are then sampled iteratively and added to the sequence, enabling autoregressive text generation. As model size and training data increase, LLMs demonstrate emergent capabilities in tasks such as reasoning, translation, summarization, and even programming~\cite{instructGPT, GPT3, GPT4, scalinglaw}. The training process typically involves several stages: large-scale unsupervised pre-training, supervised fine-tuning (SFT), reward modeling (RM), and reinforcement learning from human feedback (RLHF)~\cite{GPT4}. These stages progressively enhance the model's linguistic competence while aligning its outputs with task-specific goals.

The remarkable capabilities of LLMs have enabled diverse applications, including chatbots, code completion tools~\cite{llm_app_code}, writing assistants~\cite{llm_app_write}, and search augmentation~\cite{llm_app_search}. However, conventional LLMs remain largely confined to static, text-based interactions, which constrains their effectiveness in real-world contexts that demand interactivity and action-oriented reasoning. To address these limitations, the concept of LLM agents (\ie, autonomous AI agents powered by LLMs) has emerged, which denote systems that augment LLMs with the ability to perceive, plan and act within external environments~\cite{survey_agents, survey_agent_rise, survey_llm_tool1, survey_llm_tool2, Agent_hugginggpt, Agent_AutoGPT}.

\subsection{The Emergence of AI Agents powered by LLMs}
\begin{figure} 
	\centering
	\includegraphics[width=\linewidth]{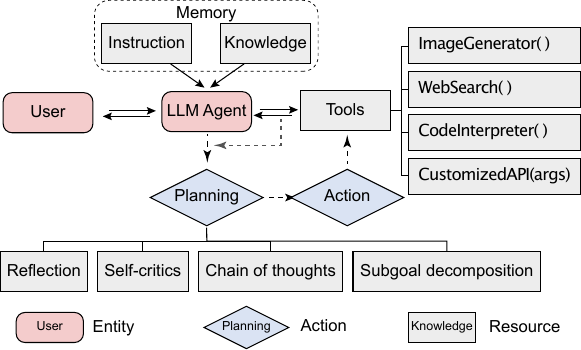}
	\caption{\textbf{The paradigm of LLM agents workflow.}
        The LLM agents are built on LLMs that incorporate integrated tools, along with memory containing pre-configured instructions and domain-specific knowledge. These agents autonomously plan and execute actions to engage with external environments in response to user inputs. Their behavior is dynamic, influenced by the user’s prompt, the model's planning preferences, pre-configured instructions, specialized knowledge, available tools, and feedback from tool executions.}
	\label{fig:agent}
\end{figure}

LLM agents represent a significant advancement over traditional LLMs by not only generating responses but also performing tasks on behalf of the user. 
These agents typically incorporate memory mechanisms to retain contextual information~\cite{Agent_memgpt}, planning modules to orchestrate action execution~\cite{survey_plan1,survey_plan2}, and interfaces for tool use, which extend their capabilities beyond language processing (Figure~\ref{fig:agent}). This architectural shift enables LLM agents to function as active problem-solvers, capable of multi-step reasoning and real-world task execution~\cite{survey_llm_tool1, survey_llm_tool2}.
By leveraging these capabilities, LLM agents can adapt their strategies dynamically, responding to feedback and evolving contexts, thus enhancing their flexibility and adaptability~\cite{Agent_react}.
Consequently, LLM agents mark a critical step toward bridging the gap between passive text generation and autonomous AI systems, with applications across various domains, including research assistance~\cite{Agent_for_study}, software development~\cite{Agent_software}, customer support~\cite{LLM_custom_support}, disease diagnosis~\cite{Agent_medical}, and personal productivity~\cite{LLM_personal_Productivity}. 

Among the various practices in online LLM agent services, \gpts{}~\cite{GPTs}, introduced by OpenAI in late 2023, stand out as a notable example of success, achieving significant success through widespread adoption and a large user base. 
These agents are highly customizable, designed to tackle domain-specific tasks across various industries. Their rapid adoption has led to millions of user-created instances, with applications ranging from customer support and technical assistance to education\cite{GPTs_creation}.
According to a snapshot on October 23, 2024, there are 755{,}297 \gpts{} created by diverse users, generating more than 172 million conversations.

Unlike the general-purpose ChatGPT (\eg, GPT-4), which is optimized for broad, open-ended interactions, \gpts{} enable end-users to create specialized agents using natural language prompts. This design lowers the barrier to agent development by allowing users to specify customized instructions without requiring programming expertise~\cite{GPTs}. Users can configure these agents with tailored instructions, distinct personality traits, and domain-specific knowledge. This flexibility supports the development of agents capable of adhering to specialized workflows, employing industry-specific terminology, and performing tasks with high precision.

\gpts{} enables end-users to create specialized agents directly using natural language prompts, thereby lowering the barrier to agent development~\cite{GPTs}. Unlike the general-purpose model of ChatGPT (\eg, GPT-4), which is geared toward broad interactions, \gpts{} allow users to configure agents with tailored instructions, distinct personality traits, and domain-specific knowledge. This flexibility fosters the development of agents capable of following specialized workflows, using industry-specific language, and performing tasks with a high degree of precision.

\begin{table}[]
\centering
\caption{\textbf{Configurations for customizing a \gpts{} and optional built-in tools/capabilities provided by OpenAI.} A \gpts{} follows the specified \textit{Instructions} and references the uploaded \textit{Knowledge} to adapt its responses. The \textit{Custom Actions} feature enables the GPT agent to interact with third-party services and perform complex tasks beyond its default capabilities.}
\label{tab:gpts-configuration}
\resizebox{\columnwidth}{!}{%
\begin{tabular}{ll}
\hline
\textbf{Configuration/Tools} & \textbf{Description}                                                            \\ \hline
Instructions        & Guidelines for functions, behavior, and constraints. \\
Knowledge           & Pre-uploaded files as additional reference context.                 \\
Browsing            & Web browsing for retrieving real-time information.                                \\
DALL·E              & Image generation.                                                      \\
Code Interpreter    & Code execution.                                                        \\
Data Analysis       & Retrieval and analysis of user-provided documents.                                              \\
Custom Actions      & Integration with 3rd-party APIs.                       \\ \hline
\end{tabular}%
}
\end{table}

To create a customized GPT, users configure three core elements: \emph{Instructions} (behavioral guidelines and constraints), \emph{Knowledge} (external files the agent can reference), and \emph{Capabilities} (extensions that enhance functionality)~\cite{GPTs_creation}. OpenAI currently provides four primary \emph{Capabilities}: (1) \emph{Actions}, which enable API calls through function call mechanisms~\cite{GPTs_actions}; (2) \emph{Browsing}, which allows real-time web access; (3) \emph{Code Interpreter \& Data Analysis}, which supports code execution and data processing; and (4) \emph{DALL·E}, which generates images.

Beyond customization, OpenAI has established an ecosystem for distribution and discovery. Users can share their agents and access \gpts created by others using unique identifiers or through the official \textit{GPTs Store}. During interactions, the underlying LLM processes user prompts, applies the agent’s pre-defined settings, invokes the required capabilities, and synthesizes the final response. Notably, all capability invocations are executed on OpenAI’s back-end servers, ensuring secure and consistent integration of external tools.

\subsection{Potential security risks of LLMs and LLM Agents}
With the rapid advancement of large language models (LLMs), concerns regarding their security and reliability have become increasingly salient. Recent research typically focuses on three primary categories: adversarial prompts, hallucinations, and harmful outputs.
\textit{Adversarial prompts} (also referred to as prompt injection or jailbreak attacks) exploit vulnerabilities in model alignment to bypass safety mechanisms~\cite{LLM_hijacking, LLM_Jailbreak, LLM_Jailbreak2, LLM_Jailbreak3}. By carefully crafting malicious inputs, attackers can manipulate the model into producing unintended outputs or leaking sensitive information. For example, an essay embedded with Adversarial instructions will cheat in scoring while reviewed by an LLM, or attackers will use Adversarial prompts to generate harmful content that violates the usage policy.
\textit{Hallucinations} refer to instances where the model generates responses that are not factually grounded or semantically accurate~\cite{LLM_hallucinations, LLM_hallucinations2}. Such outputs can propagate misinformation, reduce user trust, and cause significant downstream risks in high-stakes applications such as healthcare, law, or finance.
\textit{Harmful outputs} encompass biased, offensive, or toxic language, as well as instructions that may facilitate dangerous activities~\cite{LLM_bias, LLM_bias2}. Although alignment techniques (\eg, reinforcement learning from human feedback, content filtering) have mitigated these risks to some extent, models remain susceptible to producing unsafe outputs under certain conditions. 
Such threats primarily concern inappropriate content, including unhelpful, unsafe, or even harmful inputs and outputs, which may serve either as the source of unintentional vulnerabilities or as the target of intentional attacks. Meanwhile, when infrastructure-level attacks are excluded, the sources and victims of threats can be classified into two fundamental configurations: (1) the user as the source and the LLMs as the victim, as in cases of jailbreak or hijacking, and (2) the LLM as the source and the user as the victim, as in cases where the model generates misleading content. 

The situation becomes more complex in the context of LLM agents. The agent paradigm introduces a third party, namely the environment accessed through tool calls, which broadens both the potential sources and victims of threats. For instance, malicious users may exploit agents to perform harmful tasks through external tool calls~\cite{sec_agent_jailbreak, sec_gpts_tracker}; misconfigured or malicious agents may leak conversation content through unauthorized tool interactions~\cite{sec_agent_privacy}; and adversaries may manipulate tool call feedback by returning corrupted or adversarial values~\cite{sec_agent_dynamic_env_attack, sec_agent_dynamic_env_attack2}. Recent research has therefore emphasized risks that arise from, and are amplified by, this shift from the conventional binary user and LLM setting to a triadic user, agent, and environment setting~\cite{sec_LLM_survey}. 
However, network-level risks introduced by the LLM agent paradigm have been largely overlooked in existing research.

\subsection{Traffic Analysis}

Network traffic refers to the data exchanged during online activities, with each Internet user identified by a unique IP address. When information is transmitted, it is segmented into packets containing metadata such as the source and destination IP addresses~\cite{ip_rfc791, tcp_rfc9293}. These packets pass through a series of network devices, including Wi-Fi access points managed by local administrators, routers maintained by Internet service providers, and undersea cables subject to government monitoring, before reaching their destination. Since these intermediaries vary in their trustworthiness and can potentially observe all transmitted data, encryption has become the standard method for preserving confidentiality. In practice, plaintext is transformed into ciphertext using cryptographic algorithms and secret keys, and only the intended recipient possessing the correct key can decrypt and reconstruct the original information~\cite{tls1_3_rfc8446, https_rfc9110}.

Traffic analysis is a well-established technique in network security that enables adversaries to infer sensitive information about network activities by examining traffic characteristics, even when the payload is encrypted~\cite{wf2002}. Such inferences rely on side-channel~\cite{sec_side_channel} attributes, including packet timing, size, direction, and frequency, that remain observable despite encryption. For example, malicious activities often exhibit recognizable sequences of packet lengths that facilitate detection~\cite{wf_df}. Likewise, users employing Virtual Private Network~(VPN) or The Onion Router~(Tor) to conceal browsing behavior may still reveal visited websites, as encrypted traffic frequently retains distinctive interaction signatures~\cite{wf2011_tor, wfvpn}.

Traditionally, traffic analysis has focused on behaviors associated with specific user activities, such as malicious intrusions or web and video access~\cite{wf2002}. More recently, attention has shifted toward traffic generated by large language models (LLMs). By monitoring the exchanges between end users and LLM service providers, adversaries can infer sensitive details of user-model interactions. Prior studies have demonstrated that sequences of packet sizes can be leveraged to deduce model outputs. This arises from the fact that LLMs typically generate responses in a token-by-token manner, where each token requires a small but non-negligible computation time. To reduce latency, most services employ a streaming mechanism that transmits each token immediately once it is generated. Because tokens differ in character length, the corresponding packets also vary in size. Consequently, attackers can observe the packet-size sequence (\eg, 6, 6, 3, 2) and apply trained decoders to reconstruct the plaintext output (\eg, “\textbar Sure, \textbar there \textbar is \textbar a \textbar \ldots”)~\cite{sec_llm_prompt}.

However, existing traffic-analysis research on LLMs exhibits several limitations. First, most approaches underexploit the multi-dimensional characteristics of traffic, such as timing patterns, packet counts, and uplink flows, resulting in limited robustness. Major service providers have already adopted basic countermeasures, including batching adjacent tokens and padding transmission contents, to mitigate inference attacks based solely on packet-size sequences. These defenses are largely effective against current dialogue-reconstruction methods. Second, existing attacks predominantly target conventional text-based LLMs, overlooking the rapid emergence of multimodal interactions and agent-based architectures. As a result, current techniques fail to generalize to modern LLM services that involve richer modalities and autonomous agent behaviors.

\section{Problem Statement}
\subsection{Threat Model}

\begin{figure}[]
\centering
\includegraphics[]{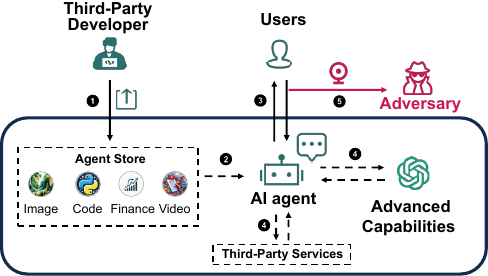}
\caption{\textbf{Detailed threat model of \ours} 
    (1) configurations of diverse LLM agents are created and subsequently made available through the Agent Store; 
    (2) an AI agent is deployed and executed on the vendor’s platform, relying on a foundation model such as GPT-4; 
    (3) a target user accesses and interacts with the designated agent through a unique identifier provided by the web service; 
    (4) the agent employs preconfigured built-in tools with advanced capabilities, together with third-party API calls, to accomplish the assigned task; 
    (5) an adversary intercepts the interaction traffic between the user and the AI agent in order to infer the ongoing task and enable further profiling of the user. 
}
\label{fig:threat-model}
\end{figure}

As illustrated in Figure~\ref{fig:main-1-overview}A~and~\ref{fig:threat-model}, we introduce a novel network-level threat on LLM agents, referred to as \ours, which exploits the network traffic generated during user-agent interactions. The objective of \ours{} is to infer the underlying agent behavior and identity, thereby revealing the information about the user.

The victim refers to a user who accesses LLM agent services via the Internet. These services usually offer a unified entry point to multiple agents, each identified by a unique ID. A user can start a conversation with a specific agent by including its identifier in the URL. For instance, \texttt{chatgpt.com/g/$<$agent\_id$>$} corresponds to a particular \gpts{} instance on ChatGPT. As illustrated in Figure~\ref{fig:main-1-overview}B~and~\ref{fig:agent}, agents may invoke specialized tools to perform tasks, process the tool outputs, and ultimately return the refined results to the user\cite{GPTs}.
In this work, we account for the diversity of user interaction paradigms by prompting the same AI agent with varied inputs, thereby better capturing realistic usage scenarios. We also adopt standard assumptions from traffic analysis~\cite{wf_df, wf_rf, wf_kfp, wf_knn}, modeling user-agent interactions as independent of other web activities and assuming communication with only a single agent at any given time.

To protect content confidentiality, users typically access agent services via HTTPS, \ie, HTTP layered on TLS encryption~\cite{https_rfc9110, tls1_3_rfc8446}, which is the standard protocol in everyday web browsing. This mechanism conceals user-agent communications from passive observers. To further reduce the risk of inference attacks based on packet size, service providers often employ defenses such as token batching and padding. Specifically, tokens generated within a short time window are combined into a single packet, which is then padded to the nearest multiple of a fixed size (\eg, 32 bytes) before transmission, thereby preventing token length from being inferred through packet size.

The attacker is modeled as a passive local adversary situated either within the same local area network (LAN) or along the end-to-end network path. This adversary can observe traffic between the user and the remote LLM service vendor but cannot modify, drop, delay, or decrypt packets. Typical examples include co-located users on a Wi-Fi network exploiting its broadcast nature, or Internet Service Providers (ISPs) and censors monitoring traffic through routers and switches. The adversary’s goal is to infer which LLM-based agent service a user is accessing, relying exclusively on traffic patterns. Because agent identities are often associated with specific functionalities, such inferences may expose sensitive user information and lead to targeted harmful or unethical actions. For instance, detecting traffic to a medical agent could enable targeted phishing attacks related to diagnosis or treatment, or facilitate discriminatory practices in medical insurance. 

We further assume that the adversary can detect the onset of user interactions with an LLM agent service provider. This can be achieved using simple techniques, such as monitoring spikes in packet frequency or mapping traffic to a known set of provider IP addresses. For instance, during our observations, \url{chatgpt.com} is consistently associated with a finite set of IP addresses over several hours. In addition, an adversary positioned near the user, such as a LAN eavesdropper, may observe additional network flows beyond those directly involving the LLM agent service provider. For example, the browser may automatically fetch third-party resources, or the user may follow redirections embedded in the LLM agent’s response. These auxiliary flows can provide further information to the adversary. To capture both cases, we evaluate two settings: \textit{primary-flow}, where the adversary only analyzes traffic between the user and the LLM agent, and \textit{mixed-flow}, where the adversary considers all network flows generated during the interaction.

Given the large number of available LLM agents, attackers typically restrict their attention to a subset of interest, namely, the monitored set. Following standard practice in traffic analysis~\cite{wf_critical, wf_df, wf_rf}, we consider two evaluation scenarios: closed- and open-world. In the closed-world scenario, the user’s target agents are limited to a monitored set, and the attacker’s task is to identify the exact agent within this set. This provides a controlled environment for benchmarking identification accuracy. The open-world scenario captures more realistic conditions where users may also interact with agents outside the monitored set. Here, the attacker must first decide whether the target agent is within the monitored set and, if so, determine its identity. 

\subsection{Problem Formulation}
As shown in Figure~\ref{fig:main-1-overview}C, \ours{} breaks down the process of revealing user privacy into two core steps. The first step is instant agent-level identification, determining which LLM agent the user is interacting with during an ongoing conversation. While this identification already discloses the immediate agent access and the likely conversation topic, this behavioral privacy leakage also establishes a foundation for further profiling. In the second step, cross-agent usage patterns (\ie, statistics of used agents over time) are utilized to infer more sensitive personal information.

Mainstream LLM services adopt encryption protocols to protect traffic payloads from interception. However, encryption does not conceal traffic metadata, including packet timing, direction, and size~\cite{https_rfc9110, tls1_3_rfc8446}. Although metadata does not expose content directly, it still encodes traffic patterns that can be associated with particular behaviors or agents.
The attack, therefore, hinges on identifying distinctive patterns within encrypted traffic generated during interactions, which can be exploited to infer both user behaviors and the identities of LLM agents.
To formalize this process, we model each user-agent interaction traffic as a traffic trace $ T $, defined as a sequence of packets exchanged between the user and the LLM agent. A trace consists of a sequence of packet tuples and is formally represented as
\begin{equation}
T = \langle P_i \rangle_{i=1}^n, \quad
\text{where } P_i = (t_i, d_i, s_i),
\label{eq:trace_representation}
\end{equation}
where $ T $ is the trace, $ P_i $ is the $ i $-th packet in the trace, $ t_i \in \mathbb{R}^+ $ is the arrival timestamp of the packet relative to the very first packet of traffic, $ d_i \in \{-1, 1\} $ indicates the direction of the packet ($-1$ for incoming and $1$ for outgoing), and $ s_i \in \mathbb{Z}^+ $ represents the packet size in bytes.

Our goal is to construct a parameterized model $ f_\theta $, implemented as a deep neural network, that maps an encrypted trace $ T $ to a classification label $ y $. Formally, this is expressed as:
\begin{equation}
    f_\theta: \mathcal{T} \rightarrow \mathcal{Y}, \quad f_\theta(T) = y,
    \label{eq:dnn_model}
\end{equation}
where $ y \in \mathcal{Y} $ is a discrete label representing a category from the predefined label space $ \mathcal{Y} $, such as the type of service behavior or the identity of the LLM agent. This formulation corresponds to a supervised multi-class classification problem, in which the model learns to infer labels from the fingerprints of encrypted traffic traces.
To enable this classification, the raw trace data must be transformed into structured numerical features. In \ours{}, we propose an effective feature extraction method that aggregates packet-level information into a time-windowed Multi-view Traffic Aggregation Matrix (MTAM), which is well-suited for capturing the fingerprints of LLM agent interaction traffic. The detailed implementation is presented in the following section.

After completing instant agent-level identification, the next step is to infer the user's occupation based on their cross-agent usage patterns over time. Ideally, we assume the adversary obtains a frequency vector $ \mathbf{f} \in \mathbb{R}_{\geq 0}^n $, where each entry $ f_i $ denotes the frequency of interaction with the $ i $th agent in a set of known LLM agents $ \mathcal{A} = \{a_1, a_2, \ldots, a_n\} $.

Under this assumption, the occupation profiling task can be formulated as a zero-shot inference problem:
\begin{equation}
    g_f: \mathbb{R}^n \rightarrow \mathbb{R}^{|\mathcal{O}|}, \quad g_f(\mathbf{f}) = \mathbf{p},
\end{equation}
where $ \mathbf{p} = [p_1, p_2, \ldots, p_{|\mathcal{O}|}] $ is a score vector that quantifies the relevance of each occupation in $ \mathcal{O} $. By ranking the entries of $ \mathbf{p} $ in descending order, the adversary obtains a prioritized list of plausible occupations rather than a single predicted label.

This formulation leverages known associations between occupations and their characteristic patterns of agent usage. The function $g$ is treated as a zero-shot inference mechanism, motivated by the absence of large-scale labeled datasets of user-agent interactions and by the need to ensure generalizability. A zero-shot approach enables inference for previously unseen users and occupations by relying on behavioral priors rather than supervised training.

In practice, however, ethical considerations and data accessibility constraints make it infeasible to obtain precise usage frequencies from real users in our experiments. By contrast, the relative ranking of agent usage, that is, identifying which agents are used more frequently than others, can be more feasibly estimated through user surveys or simulated sampling. Accordingly, our experimental design adopts a rank-based representation:
\begin{equation}
    \mathbf{r} = \operatorname{rank}(\mathbf{f}), \mathbf{r} \in \mathbb{Z}^n,
\end{equation}
where $ \mathbf{r} $ is a permutation indicating the descending order of agent usage frequency. 

Meanwhile, since the input cross-agent usage is derived from the results of agent-level classification, it is inherently affected by the classification error in the previous step. To capture this uncertainty, we model the observed rank vector as a noisy permutation $ \tilde{\mathbf{r}} $, where each rank position has a certain probability of being randomly swapped with another. This simulates the impact of agent identification errors on downstream profiling.

Accordingly, the final occupation profiling task is formalized as:
\begin{equation}
    g_r: \mathbb{Z}^n \rightarrow \mathbb{R}^{|\mathcal{O}|}, \quad g_r(\tilde{\mathbf{r}}) = \mathbf{p},
    \label{eq:rank_based_occupation_inference}
\end{equation}
where $ \tilde{\mathbf{r}} $ is the perturbed agent rank vector and $ \mathbf{p} $ denotes the corresponding occupation relevance scores. This formulation enables zero-shot inference based solely on the noisy ordering of agent usage, making the model robust to prior classification errors and more practical in realistic attack scenarios.
In the next section, we describe how these rank-based features are used for zero-shot occupation inference.

\section{Design of \ours}
In this section, we introduce the core methods underpinning \ours. As illustrated in Figure~\ref{fig:main-1-overview}A, \ours{} reveals LLM agent users' privacy by first identifying the ongoing interaction objective and then profiling user attributes through cross-agent usage patterns over time. 
To realize \ours, we first propose a strategy for generating diverse and tailored prompts for distinct agents by leveraging both external observation and internal introspection, thereby facilitating high-quality interaction data collection (see Figure~\ref{fig:main-3-design}A). 
We then introduce a multi-view feature extraction method that exploits the \textit{processuality} and \textit{multimodality} of the traffic, and train a CNN-based classification model for accurate agent identification (see Figure~\ref{fig:main-3-design}B). 
Finally, we develop a zero-shot occupation inference method that represents each agent through its alignment with Detailed Work Activities (DWAs). These representations are embedded into a DWA-based occupation network, which utilizes modularity variations to compute relevance scores and infer the user’s occupation type (see Figure~\ref{fig:main-3-design}C).

\subsection{Diverse Tailored Functional Prompt Generation}
To effectively utilize traffic fingerprints, a common practice in traffic analysis~\cite{wf_cumul, wf_kfp, wf_knn, wf_df} is to collect corresponding traffic repeatedly for each specific target class (\eg, websites~\cite{wf_df, wf_cumul, wf_holmes, wf_ares}, video segments~\cite{wf_video}, or malicious payloads~\cite{}). This process allows the model to learn traffic patterns specific to each class. However, designing an effective data collection strategy for \ours{} presents several challenges.

First, compared to conventional traffic analysis scenarios, user-agent interaction traffic exhibits both inter-class similarity and intra-class diversity, complicating the extraction of discriminative features. Inter-class similarity arises because different AI agents share the same front-end entry point on the service platform, leading to similar global traffic patterns due to the identical sequence of shared web components during loading. This reduces variance across classes. In contrast, intra-class diversity stems from the high variability in conversation content, even within the same agent, driven by diverse user inputs and the inherent randomness of LLM-generated responses.
Second, data collection must align with the goals of downstream analysis. Since \ours{} aims to reveal user information by analyzing traffic associated with specific AI agents, it is crucial to capture interactions that reflect each agent's core functionality. Given the broad capabilities of LLM-based AI agents, not all queries will elicit responses that reflect their specialized roles. For example, an email-writing agent may respond to a healthcare-related query using only the base model’s generic language abilities, without invoking specialized tools for composing or sending emails. In such cases, the response remains plain text, offering limited insight into the agent's distinctive behavior and contributing little to downstream profiling via cross-agent usage. Therefore, it is essential to ensure that the collected data captures representative functionality.
Finally, large-scale random data collection is impractical due to cost and access limitations. Unlike ordinary applications, LLM-based AI agents are subject to usage constraints and monetization limits. For example, OpenAI's \gpts{} currently imposes a rate limit of 40 requests per 3 hours during our experiments, significantly restricting the throughput of interaction data acquisition.

Given that an agent's behavior is driven by the instructional prompt, the quality of the collected traffic data heavily depends on the quality of these prompts used. Accordingly, we define the design objective as generating prompts that are \textit{Diverse}, \textit{Tailored}, and \textit{Functional}.
Diverse prompts ensure that each interaction reflects a distinct scenario, covering a broad spectrum of the agent's capabilities and simulating realistic, varied user inputs.
Tailored prompts are customized for each specific agent, targeting its unique functional domain and behavioral traits.
Functional prompts are designed to reliably activate the agent's core functionalities, ensuring that the interactions capture the distinctive behavioral patterns necessary for effective downstream analysis.

To this end, we propose a prompt generation framework that integrates both external observation and internal introspection to generate diverse, tailored, and functionally meaningful prompts for each agent. This approach facilitates the collection of high-quality interaction traffic data. As illustrated in Figure~\ref{fig:main-2-observation}B, we combine external behavioral analysis with intrinsic agent self-reflection to generate prompts that elicit representative behaviors.
The process begins by employing self-instructional techniques to prompt the agent to generate diverse prototype queries that reflect its core functionalities. Next, we gather publicly available agent attributes and conduct manual conversational case studies to summarize each agent's typical capabilities and behavioral patterns. This summary is then used to refine and constrain the self-instructed prompts. The refined set of diverse, tailored prompts consistently activates the agent's characteristic functionalities. These prompts are subsequently employed to simulate realistic interactions, thus enabling the collection of high-quality traffic data for \ours{} and facilitating downstream user profiling.

We begin by employing a self-instruction technique to generate prototype prompts, leveraging the agent's internal introspection and generative capabilities. Specifically, we initiate a multi-turn dialogue with the agent, beginning by asking it to reflect on its core functionalities. We then prompt the agent to envision several realistic user scenarios where each function might be needed and finally request it to generate user-like prompts that would effectively activate the corresponding functionalities.
This approach capitalizes on the agent's internal introspection, \ie, its access to hidden, pre-configured knowledge and instructions that are typically inaccessible to external entities. As noted earlier, LLM-based agents are built upon domain-specific configurations, but these configurations are often not fully documented and may be obscured by generic names and brief descriptions. For example, while an AI agent named \textit{Diagram} is clearly intended for diagram generation, its precise capabilities are not immediately apparent from public metadata alone.
Through introspective querying, \textit{Diagram} revealed six specific diagram types it can generate: \textit{flowchart}, \textit{mind map}, \textit{sequence diagram}, \textit{Sankey diagram}, \textit{class diagram}, and \textit{timeline}. Based on this insight, we instructed the agent to generate diverse prompt examples for each type. This process resulted in a collection of 150 high-quality prompts, covering a broad range of realistic user needs across the agent's functional scope.

Building on the prototype self-instructed prompts, we further refine them through a combination of public metadata analysis and manual conversational case studies. We begin by examining available attributes such as the agent's name, description, example starters, developer identity, and the list of enabled external tools. These details help us infer the agent's functional domain, interaction patterns, and tool invocation behavior, thereby supplementing the self-instructed prompts with more precise and explicit functional triggers.
Simultaneously, we conduct manual studies of example conversations to extract common input requirements and behavioral patterns that may not be apparent from the metadata alone. For instance, the agent \textit{AskYourPDF Research Assistant} requires users to upload a PDF file; \textit{Tattoo GPT} expects structured inputs specifying theme, style, placement, size, and color; and \textit{Character GPT} necessitates a multi-turn dialogue with explicit confirmation before generating an image. These insights are then used either to directly refine the prototype prompts or to improve the self-instruction process itself, resulting in high-quality prompts that reliably elicit the agent's distinctive functionalities.
The final data collection settings, which incorporate these refinements, are outlined in the following section.

\subsection{Multi-view Interaction Traffic Extraction and Classification}
As previously defined, each raw trace consists of a sequence of packets characterized by timestamps, directions, and sizes. However, due to the sparsity and variable length of these traces, they cannot be directly input into standard machine learning models. To effectively capture the behavioral patterns embedded in encrypted traffic traces, we adopt and extend the Traffic Aggregation Matrix (TAM), a technique proposed in prior research on encrypted traffic classification~\cite{wf_rf}. This method transforms each variable-length trace into a fixed-size, structured feature matrix.

Specifically, we divide the duration of each trace into a fixed number of temporal windows. In our experimental setup, each trace is uniformly partitioned into $ W = 1800 $ non-overlapping segments of equal width. Within each time window $ w_j $, we extract statistical features separately for inbound (\textit{in}) and outbound (\textit{out}) traffic directions. For each direction $ d \in {in, out} $, we compute two types of statistics:
(1) The number of packets, $ N_d(j) $, and
(2) The total transmitted data volume, $ B_d(j) $, in window $ w_j $ for direction $ d $. These statistics are defined as follows:
\begin{equation}
N_d(j) = \sum_{i=1}^{n} \mathbb{I}(t_i \in w_j \land d_i = d),
\end{equation}
and
\begin{equation}
B_d(j) = \sum_{i=1}^{n} \mathbb{I}(t_i \in w_j \land d_i = d) \cdot s_i,
\end{equation}
where $ n $ is the total number of packets in the trace, $ t_i $ is the timestamp of packet $ i $, $ d_i $ is its direction, and $ s_i $ is its size in bytes. The indicator function $ \mathbb{I}(\cdot) $ evaluates whether the packet falls within the time window $ w_j $ and matches the specified direction $ d $, returning 1 when the condition is met and 0 otherwise. Thus, $ N_d(j) $ counts how many packets are sent or received during that window, while $ B_d(j) $ sums the total bytes transmitted in that direction.

Notably, the number of packets and the total transmitted bytes are not linearly correlated (see Figure~\ref{fig:main-2-observation}), as packet sizes can vary significantly. By incorporating both statistics, we aim to capture the fingerprint of the interaction traffic: the packet counts reflect the structural and temporal dynamics driven by process workflows (\eg, information exchange latency and sequence), while the transmission volumes reflect the variability introduced by multimodal outputs (\eg, types and contents of payloads).

Finally, each trace is transformed into a Multi-view Traffic Aggregation Matrix (MTAM), with a shape of $ 2 \times 2 \times W $: two channels corresponding to the packet count $ N $ and byte volume $ B $. Each channel contains directional information for both inbound and outbound traffic. This representation preserves the temporal structure of the trace and embeds rich traffic features in a consistent format, forming a robust foundation for downstream model training.

We employ a convolutional neural network (CNN) to extract fingerprints from the previously constructed MTAMs. Our design is based on the classifier used in RF~\cite{wf_rf}, with the key modification of extending the input from a single channel to a dual-channel structure, incorporating both packet count and byte volume.
The input is first passed through two 2D convolutional blocks. Each block consists of two convolutional layers, followed by ReLU activation, batch normalization, max pooling, and dropout. These 2D convolutions jointly capture temporal and directional patterns in the input matrix. The resulting feature map is then reshaped into a 1D sequence, and the number of channels is reduced to 32, preparing the data for the subsequent 1D convolutional layers.
Next, two 1D convolutional blocks are applied to further model the temporal evolution of traffic behaviors and extract higher-level features. The output is then passed through a 1D convolutional layer, where the number of output channels equals the number of classes, followed by a global average pooling (GAP) layer. Finally, a softmax layer outputs the classification probabilities.
This architecture effectively balances parameter efficiency with the ability to model rich behavioral patterns by integrating bidirectional and multi-dimensional traffic statistics, which are critical for accurate identification in downstream tasks.

\subsection{Zero-shot User Attribute Profiling}
Building on the identification of user-agent interaction traffic, the next stage of \ours{} aims to profile sensitive user attributes by exploiting cross-agent usage patterns. Such profiling can facilitate downstream attacks, including spear-phishing campaigns tailored to the user’s personal context.
In this paper, we use occupation type as a prototype for assessing the risks associated with user information exposure. The same approach could be extended to profile other categories of personal attributes.

We focus on occupation profiling for three reasons. First, LLM-based AI agents are often designed for domain-specific professional functions and are widely employed in professional tasks, making cross-agent usage behavior a strong indicator of occupational roles. Second, occupations constitute high-value profiling targets from a security perspective, as they provide actionable contexts for threats such as employment scams and professional social engineering. Third, the inference task is challenging because the relationship between agents and occupations is many-to-many and often indirect.
To address these challenges, we introduce a zero-shot occupation profiling method that infers occupational categories from user-agent interaction patterns without relying on labeled user data.

We begin by representing each agent in terms of the set of Detailed Work Activities (DWAs) it supports. DWAs, as defined in the O*NET database~\cite{onet2023}, are fine-grained descriptions of work activities that decompose the tasks associated with specific occupations. O*NET defines 923 standard occupations, 4{,}712 tasks, and 21{,}458 DWAs. Each occupation $ o \in \mathcal{O} $ is composed of a set of tasks $ \mathcal{T}_o \subseteq \mathcal{T} $, and different occupations do not share tasks.  
Each task $ t \in \mathcal{T} $ is further associated with a set of DWAs $ \mathcal{D}_t \subseteq \mathcal{D} $, which are often shared across tasks.
The DWA profile for each occupation is defined as:
\begin{equation}
\mathcal{D}_o = \bigcup_{t \in \mathcal{T}_o} \mathcal{D}_t.
\end{equation}

Based on the insight that AI agents assist professionals in performing DWAs, we associated each agent $ a \in \mathcal{A} $ with a set of DWAs $ \mathcal{D}_a \subseteq \mathcal{D} $, representing the work activities that the agent can facilitate or automate. 

The similarity between an agent's DWA set and that of an occupation is measured using the Sørensen-Dice coefficient~\cite{sørensen1, sørensen2}:
\begin{equation}
\text{Sim}_{\text{Sørensen}}(\mathcal{D}_a, \mathcal{D}_o) = \frac{2 \cdot |\mathcal{D}_a \cap \mathcal{D}_o|}{|\mathcal{D}_a| + |\mathcal{D}_o|}.
\end{equation}

To model the semantic relationships between occupations, we construct an occupation graph as proposed by~\cite{GPT_labor_science}. Each node corresponds to a detailed occupation, and edge weights reflect the Sørensen similarity between their DWA profiles.  
Following~\cite{GPT_labor_science}, we apply the Louvain algorithm~\cite{louvain} to the occupation graph to split the nodes (\ie, detailed occupations) into several communities, and adopt the resulting $ K = 12 $ communities and their labels for consistency and comparability. For the weighted occupation network $ \mathcal{G} $ with a given community division $ \mathcal{C} $, the modularity $ Q_{\mathcal{G}}(\mathcal{C}) $ measures the strength of the community structure, which is defined as:
\begin{equation}
\label{eq:modularity}
Q_{\mathcal{G}}(\mathcal{C}) = \frac{1}{2m} \sum_{i,j} \left[ A_{ij} - \frac{s_i s_j}{2m} \right] \delta(c_i, c_j),
\end{equation}
where $ A_{ij} = \text{Sim}_{\text{Sørensen}}(\mathcal{D}_{o_i}, \mathcal{D}_{o_j}) $ denotes the edge weight between nodes $ i $ and $ j $,  
$ s_i = \sum_j A_{ij} $ is the strength of node $ i $,  
$ m = \frac{1}{2} \sum_{i,j} A_{ij} $ is the total weight of all edges in the graph,  
$ c_i $ denotes the community assignment of node $ i $, and  
$ \delta(c_i, c_j) = 1 $ if nodes $ i $ and $ j $ belong to the same community, and 0 otherwise.

To estimate the affinity between an agent and each community, a naive approach might average the similarity scores between the agent and occupations within each community, or alternatively, insert the agent into the graph and recompute the community structure. However, the former fails to fully exploit the underlying semantic structure of the network, while the latter introduces structural inconsistencies, as agents and occupations differ semantically, and adding external nodes may distort the intrinsic network topology.

Instead, we propose a probing-based method grounded in modularity variation. Our probing approach allows us to preserve the original community structure while assessing an agent's potential integration.
We treat the agent as a virtual node and hypothetically insert it into each existing community while keeping the original community structure fixed.  
For each candidate community $ c_k $, we compute the modularity of the augmented graph $ \mathcal{G}_a = \mathcal{G} \cup \{a\} $ under the fixed partition $ \mathcal{C}_k $, denoted as $ Q_{\mathcal{G}_a}(\mathcal{C}_k) $, using Equation~\ref{eq:modularity}.  
This process simulates probing the agent node into the $ k $-th community without altering the community assignments of other nodes.

We define the modularity variation as:
\begin{equation}
\begin{aligned}
\Delta Q_{\mathcal{G}_a}^{(k)} &= Q_{\mathcal{G}_a}(\mathcal{C}_k) - Q_{\mathcal{G}_a}(\mathcal{C}_{K+1}) \\ &= \frac{1}{m'} \left( \sum_{i \in \mathcal{C}_k} A_{ia} - \frac{s_a s_{\mathcal{C}_k}}{2m'} \right),
\end{aligned}
\end{equation}
where $ \mathcal{G}_a = \mathcal{G} \cup \{a\} $ is the augmented graph with the agent node $ a $,  
$ m' $ is the total weight of edges in $ \mathcal{G}_a $,  
$ s_a = \sum_i A_{ai} $ is the strength of node $ a $, and  
$ s_{\mathcal{C}_k} = \sum_{i \in \mathcal{C}_k} s_i $ is the total strength of community $ \mathcal{C}_k $.  

Compared to the naive approach of simply averaging similarity scores, this metric captures both connection density and degree distribution, providing a more principled and structurally informed measure of community affinity at the agent level.

To aggregate multiple agent-level signals into a user-level prediction, we aim to capture both the relative functional advantage of each agent and its overall discriminative value with respect to the occupational landscape.  
The former emphasizes agents that are relatively more aligned with a given community compared to others (\eg, if several agents are highly aligned with community $ c_1 $, the one with the weakest alignment is treated as less informative).  
The latter accounts for agents that exhibit non-specific or uniformly low affinity across all communities, such as a religion-related agent with average modularity scores, which provides limited value for occupation inference.
To this end, we compute the relative modularity variation by applying the Revealed Comparative Advantage (RCA) framework to the modularity-based affinity matrix. The RCA score is computed as:
\begin{equation}
\text{RCA}_{a,c} =  \frac{Q_c(a) / \sum_{c'} Q_{c'}(a)}{\sum_{a'} Q_c(a') / \sum_{a',c'} Q_{c'}(a')},
\end{equation}
which reflects the extent to which agent $ a $ is associated with community $ c $, relative to other agents and communities.
For simplicity, we denote the modularity-based affinity between a specific agent $ a $ and community $ c $ as $ Q_c(a) $, which corresponds to $ Q_{\mathcal{G}_a}(\mathcal{C}_c) $ as defined above. 
An RCA score less than 1 indicates a weaker association, while a score greater than 1 suggests that the agent is relatively more strongly aligned with the community. 
Additionally, to improve interpretability and enable linear scaling, we apply a logarithmic transformation to the RCA score to obtain the final agent-community association score 
\begin{equation}
\text{R}_{a,c} = \log{\text{RCA}_{a,c}}.
\end{equation}

To infer the user's occupation, we aggregate RCA scores across the cross-agent usage of the user.  
Since exact usage frequencies are unavailable in practice, we adopt an Exponentially Weighted Moving Average (EWMA) over the ranked agent list:
\begin{equation}
\hat{y}_c = \sum_{i=1}^{|\mathcal{A}_u|} \alpha (1 - \alpha)^{i-1} \cdot \text{R}_{a_i,c},
\end{equation}
where $ \alpha \in (0, 1) $ is the decay factor and $ a_i $ denotes the $ i $-th most important agent.
The predicted occupation corresponds to the community $ c $ with the highest aggregated score $ \hat{y}_c $.

\section{Evaluation}
In this section, we provide a systematic evaluation of \ours{} through experiments on agent traffic fingerprinting and user attribute inference. We describe the datasets used, including self-collected user-agent interaction data and public datasets containing information on user attributes and LLM agents. The evaluation focuses on identifying the behaviors and identities of LLM agents, analyzing user attribute exposure and privacy risks, and profiling occupations based on agent identification results. The findings highlight the effectiveness of our approach in distinguishing user-agent interaction fingerprints and assessing privacy risks, particularly regarding sensitive user attributes like occupation.
\subsection{Experimental Setup}
We begin by introducing the datasets used in our experiments of traffic fingerprinting and user attributes inference experiments.
We conduct the agent behavior and identity fingerprinting experiments on a self-collected user-agent interaction traffic dataset. The dataset contains 7{,}311 traces corresponding to 50 top \gpts{}~\cite{top500gpts} as monitored agents with distinct prompts for each class and 428 traces from 166 unmonitored agents. 
We also conduct fingerprinting privacy risk analysis on 12{,}432 \gpts{} with more than 1{,}000 historical conversations according to the snapshot~\cite{sec_gpts_tracker} on October 23, 2024.
For occupation inference, we adopt the O*NET database~\cite{onet2023} as the reference taxonomy of occupational categories. 
We construct 5{,}538 virtual user profiles on LLM agents, and also conduct a real-world user study involving 49 participants to collect their actual agent usage preferences.

The identification model is implemented in PyTorch and trained on a single NVIDIA RTX 4090 GPU. 
For each agent, the dataset is randomly split into training, validation, and test sets with a ratio of 8:1:1. 
To mitigate random fluctuations, we perform 10-fold cross-validation, randomly partitioning the dataset into 10 distinct train-test splits. 
In each fold, the model is trained and validated on the training set, evaluated on the test set, and the reported performance is averaged across all folds.
We use the macro-averaged F1 score as the primary evaluation metric to evaluate the classification performance of the fingerprinting model, as it assigns equal weight to each class regardless of its frequency, thus providing a balanced measure of performance in our multi-class classification setting. We use top-$K$ accuracy to assess the occupation profiling effectiveness, where a prediction is considered correct if the true label appears in the top $K$ outputs.

\subsection{Interaction Traffic Collection}

To collect the traffic data required for \ours, we first describe the criteria for selecting the LLM agents used in our experiments, then outline the prompt generation process, and finally detail the procedures for data collection, filtering of low-quality samples, and analysis of the resulting data distribution.

To ensure evaluation in a realistic and representative setting, we adopt \gpts~\cite{GPTs}, the agent services provided by OpenAI, as the concrete instantiation of LLM agents in our study. In comparison with other platforms, such as \textit{Wenxin AgentBuilder}~\footnote{\url{https://agents.baidu.com/center}} and \textit{Tongyi Agent}~\footnote{\url{https://www.tongyi.com/discover}}, \gpts{} are particularly suitable because they are among the earliest to be deployed, provide broad functionality across domains, and maintain a large, active user base~\cite{sec_gpts_tracker}. These features make \gpts{} a practical and representative choice for examining real-world patterns of agent usage. 
The agents for interaction traffic collection are selected from an open-source repository~\cite{top500gpts}, which publishes a daily updated ranking of the top 500 \gpts{} based on the number of recorded user interactions in the official \gpts{} store. This dataset provides a transparent and data-driven basis for identifying widely adopted agents with diverse functionalities.

\begin{figure} 
	\centering
	\includegraphics[width=\linewidth]{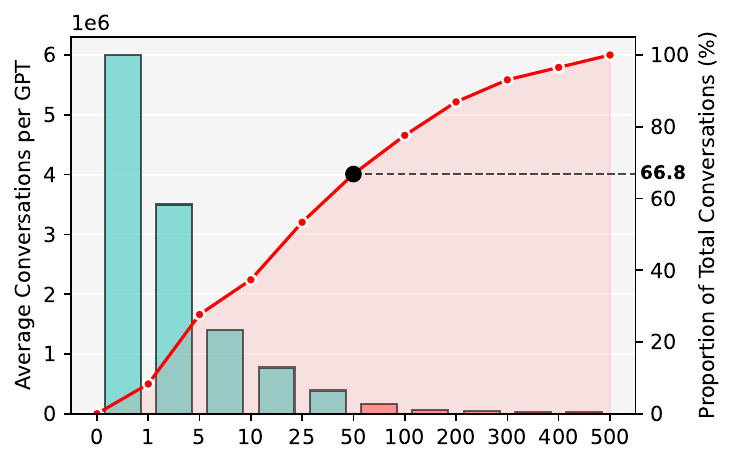}
	\caption{\textbf{Long-tailed distribution of historical conversation numbers for the top 500 most popular GPTs as of August 30, 2024.}
        Bars show the average conversations per GPT by rank interval (left axis), with teal for top-50 and red for ranks 51-500. The red line (right axis) indicates the cumulative percentage of total conversations. The vertical dashed line marks the top-50 threshold, showing that the top 50 GPTs account for 66.8\% of all conversations.}
	\label{fig:gpts_dist}
\end{figure}

As shown in Figure~\ref{fig:gpts_dist}, the conversation counts of \gpts{} follow a long-tail distribution, with a small number of highly popular agents accounting for a disproportionately large share of interactions. To ensure both practicality and representativeness, we select the top 50 \gpts{} for conducting \ours{} and experimental validation. Together, these agents account for 48 million visits, representing 66.67\% of all visits to the top 500 \gpts{} as of August 30, 2024. According to the official classification, the selected agents cover a diverse range of domains, including \textit{Writing}, \textit{Productivity}, \textit{Research \& Analysis}, \textit{Lifestyle}, and \textit{Programming}. 
During this selection process, we manually verify and exclude certain \gpts{} based on predefined criteria. In particular, agents offering identical functionalities built on the same backend services are treated as duplicates. For example, \textit{AI Humanizer} and \textit{Humanize AI} both provide text humanization and are operated by the same provider, \textit{gptinf.com}.
Furthermore, since our goal is to identify ongoing agent behaviors and uncover potential privacy risks associated with LLM-based agent services, we exclude \gpts{} that do not employ any specific agent capabilities. For example, \textit{Language Teacher \textbar Ms. Smith} operates similarly to the base model and is therefore excluded. To ensure relevance, all selected \gpts{} are required to incorporate at least one of the following extended capabilities: \textit{Action} (interaction with third-party APIs), \textit{Analysis} (document analysis), \textit{Browsing} (access to web content), \textit{Interpretation} (code execution), \textit{Multimodal Generation} (generation files, images, videos, or audio), or \textit{Redirection} (embedding third-party URLs within responses for external navigation).

We adopt the previously introduced prompt generation method to construct diverse, tailored, and functional prompts for downstream interaction data collection. To enable realistic evaluation while mitigating the risk of artificial information leakage, we employ several prompt design strategies aimed at avoiding artificially simplified tasks and instead increasing task difficulty:
First, we control the randomness of prompt length to reduce potential leakage through input size. For agents with similar functionalities, we deliberately apply identical prompts to introduce ambiguity and thereby increase the challenge of identification. In addition, we prioritize the generation of informative prompts that elicit single-turn interactions whenever possible. In contrast to multi-turn dialogues, which can often be broken down into a series of semantically independent single-turn tasks that collectively reveal more information, single-turn interactions are atomic in nature and therefore inherently more difficult to identify. Finally, for agents requiring external inputs such as PDF documents, images, or video URLs, we manually collect diverse and meaningful random inputs to ensure sufficient variability. 
In total, we generate 150 prompts per agent to support comprehensive data collection.

Next, we develop an agent interaction traffic collection framework using \textsc{Selenium}~\cite{selenium} to simulate interactions between users and agents, and employ \textsc{Tcpdump} to capture the network traffic generated by the browser during these interactions. The framework is deployed on three Ubuntu 22.04 hosts, each with 2 CPUs and 4 GB of memory, located in Singapore on Alibaba Cloud. 
Before initiating an interaction, agents requiring authentication are randomly assigned an authentication resistance status to simulate heterogeneous user environments. 
At runtime, the framework randomly selects a target agent and accesses it through the unified entry point. It then waits for a short random interval for the page to load, inputs one of the designed prompts, uploads any required files, and submits the request. The framework subsequently waits for the agent’s responses to be generated and displayed. 
During this process, the framework monitors the response process, automatically clicking the permission button if authentication is required. It also sends multi-turn prompts when an agent expects specific interaction patterns. If redirect links appear in the response, the framework follows one with a predefined probability of $p=0.5$ to simulate realistic user behavior.

To ensure the quality of the collected data, we filter out low-quality samples that do not meet our criteria. Specifically, we retain only those conversations that either trigger at least one of the agent's external capabilities or have sufficient length for a meaningful interaction. In practice, we discard text-only samples with fewer than 500 characters, as these are typically too brief to engage the agent’s functionality. Ultimately, we obtain 11{,}287 traces from 50 monitored \gpts{} and 166 unmonitored \gpts{}. Of these, 7{,}311 traces correspond to distinct prompts for each monitored \gpts{}.

To better contextualize our experiments, we analyze the distributional characteristics of the self-collected user-agent interaction dataset from three perspectives: the occurrence and co-occurrence of external agent capabilities during conversations, and the text length of the responses.

\begin{figure} 
	\centering
	\includegraphics[width=\linewidth]{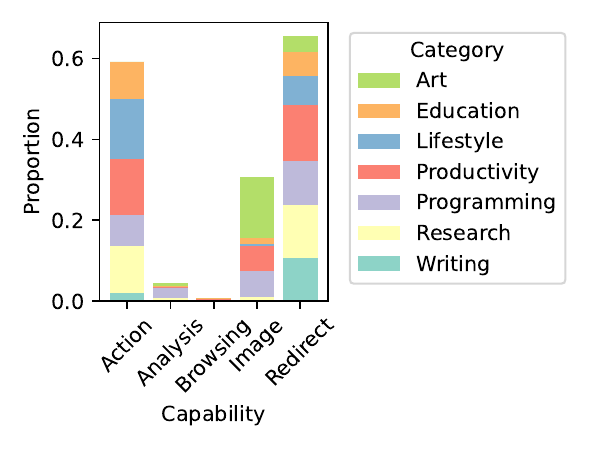}
	\caption{\textbf{Proportion of samples triggering specific capabilities across agent categories in the closed-world dataset.}
        For each agent category, sample counts are normalized to allow comparability across categories.
        Stacked bars represent the proportions of samples triggering each capability, with colors denoting different categories.
        The distribution indicates that agent categories differ substantially in the frequency with which they trigger specific capabilities.
        }
	\label{fig:sample_dist_capability}
\end{figure}

We first examine the distribution of triggered capabilities across agent categories (Figure~\ref{fig:sample_dist_capability}). Each sample is labeled with the capability it activates (\textit{Action}, \textit{Analysis}, \textit{Browsing}, \textit{Image}, or \textit{Redirect}), and the proportions are normalized within each category for comparability.
In this context, \textit{Redirect} refers to instances where the agent directs the user to an external resource via a hyperlink, indirectly reflecting the agent's functionality through subsequent user interactions, rather than through direct tool invocation.

The distribution reveals an imbalance in the combination of capabilities and agent categories. For example, \textit{Art} agents are far more likely to demonstrate image generation capabilities than to invoke third-party APIs, whereas \textit{Code} agents primarily execute code. These preferences align with common assumptions. The imbalance underscores the correlation between agent behavior and category, offering insights into how tool invocation patterns can reveal the topics of user conversations and potentially contribute to downstream user profiling.

\begin{figure*} 
	\centering
	\includegraphics[width=\linewidth]{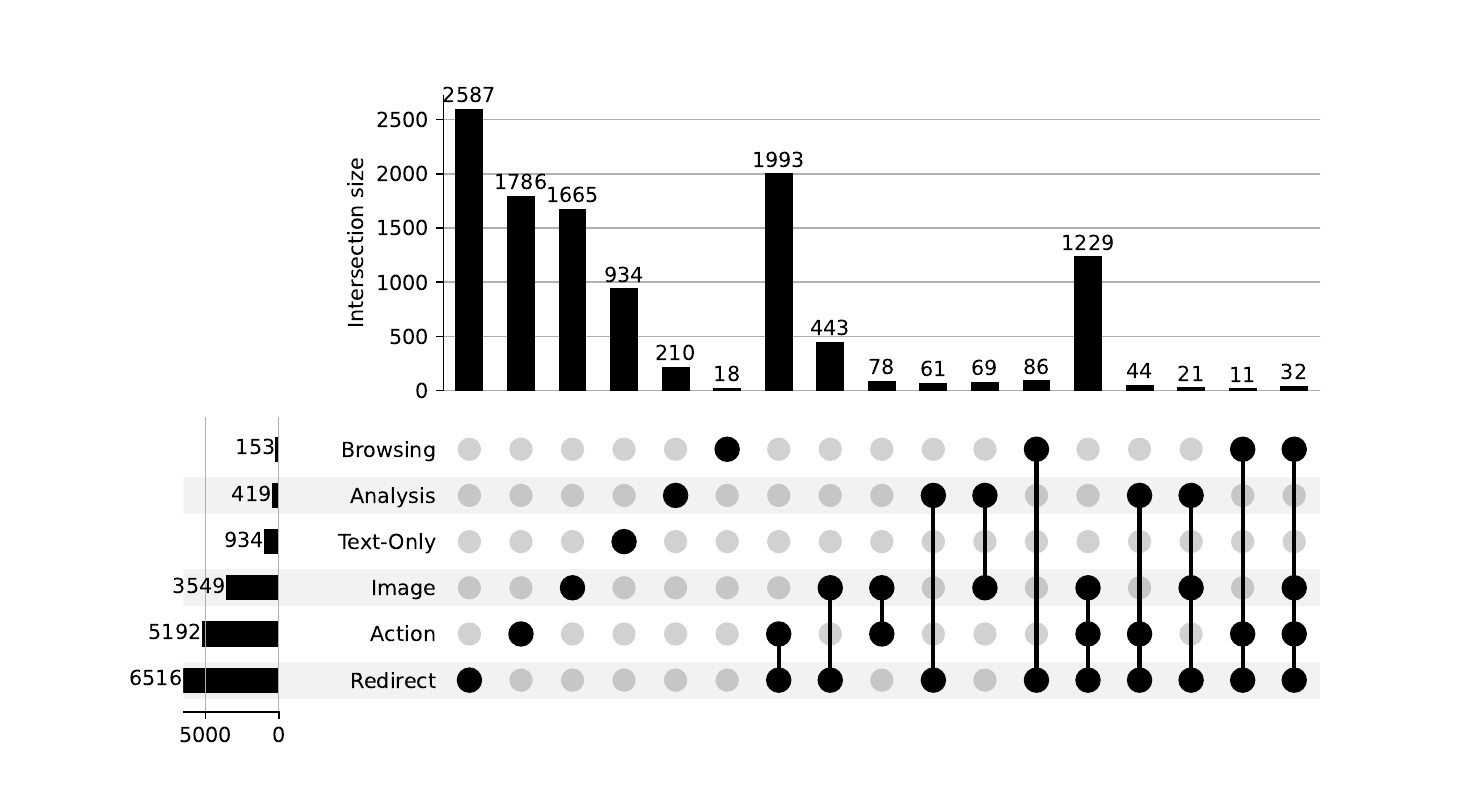}
	\caption{\textbf{Distribution of triggered capability combinations in the full dataset.}
        The main bar chart shows the number of conversations triggering different combinations of capabilities within a single interaction.
        The sidebar chart summarizes the total number of conversations triggering each individual capability.
        Results reveal substantial variation in the frequency and co-occurrence patterns of triggered capabilities.
        }
	\label{fig:sample_dist_upset}
\end{figure*}

We next analyze the co-occurrence patterns of capabilities using an UpSet plot (Figure~\ref{fig:sample_dist_upset}). Some capabilities appear independently, whereas others, most notably \textit{Action} and \textit{Redirect}, frequently occur together within the same conversation. These patterns indicate that real-world agent usage often relies on multi-step or multi-tool workflows, which may shape the similarity of traffic signatures across behavioral categories.

\begin{figure} 
	\centering
	\includegraphics[width=\linewidth]{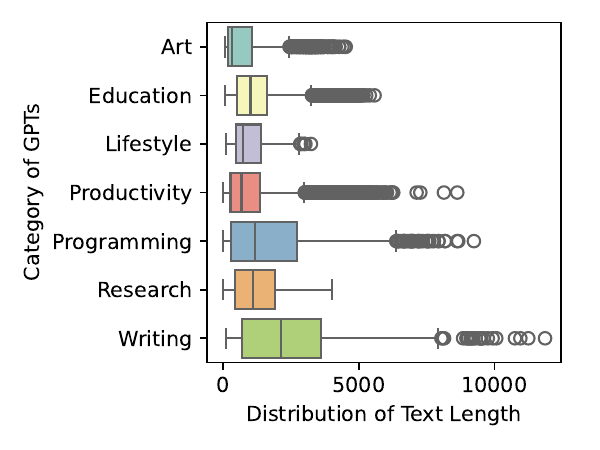}
	\caption{\textbf{Distribution of text response lengths across agent categories in the closed-world dataset.}
        Length is measured in characters.
        Boxplots show the median, quartiles, and potential outliers for each category.
        Response length distributions vary across categories, with \textit{Writing}, \textit{Programming}, and \textit{Productivity} agents producing longer outputs on average.
        }
	\label{fig:sample_dist_textlen}
\end{figure}

We then examine the distribution of text response lengths (Figure~\ref{fig:sample_dist_textlen}), measured in characters, to capture content characteristics not directly tied to specific capabilities. The boxplots reveal considerable variation across agent categories: \textit{Writing}, \textit{Programming}, and \textit{Productivity} agents generally produce much longer outputs, providing more elaborate content, whereas categories such as \textit{Lifestyle} and \textit{Art} tend to yield shorter responses. Such differences in textual verbosity may indirectly influence the volume of generated traffic and, consequently, the distinguishability of particular behaviors or agents.

\subsection{Distinguishing LLM Agent Behaviors}

We first demonstrate the feasibility of \ours{} for distinguishing LLM agent behaviors through user-agent traffic patterns. As outlined in the Literature Review, LLM agents operate under predefined instructions and dynamically invoke tools to address user requests. The final response is constructed by integrating the outputs of these tools. Accordingly, the selection and invocation of different tools subtly influence observable traffic characteristics, including timing and payload size.

We train a classifier on our self-collected user-agent interaction dataset under mixed-flow setting to identify five behavioral categories: \textit{Action}, \textit{Analysis}, \textit{Image}, \textit{Redirect}, and \textit{Plain Text}. The \textit{Browsing} category is excluded due to its frequent co-occurrence with other behaviors. We include the \textit{Redirect} and \textit{Plain Text} categories, which capture supplementary agent behaviors. \textit{Redirect} reflects agent activity indirectly through user clicks, whereas \textit{Plain Text} corresponds to purely internal responses without tool invocation. 

As shown in Figure~\ref{fig:main-4-traffic_result}B, the classifier achieves a macro-averaged F1 score of 88.9\%, demonstrating that distinct agent behaviors can be reliably distinguished using only network traffic data. 
Most categories are well separated; however, some misclassifications occur between \textit{Redirect} and \textit{Plain Text}. Specifically, 10 \textit{Plain Text} samples are misclassified as \textit{Redirect}, and 7 \textit{Redirect} samples as \textit{Plain Text}. Moreover, 11 \textit{Redirect} samples are mistaken for \textit{Action}, suggesting that certain redirection behaviors may resemble more active tool-invocation patterns.

These results confirm that the LLM agent's behaviors during user-agent interactions can be effectively fingerprinted and identified through network traffic analysis, even without inspecting message content. The strong classification performance supports our hypothesis that behavior-level distinctions in AI agents manifest as observable and distinguishable network-level patterns.

We further visualize the evolution of behavior-related features across the model’s stages using t-SNE (Figure~\ref{fig:main-4-traffic_result}A). At the input level, samples from different behavioral categories are heavily intermingled, indicating limited separability in the raw traffic features. As the model processes the data, the clusters progressively become more distinct, and by the final layer, the samples form compact groups that align closely with the behavioral categories. This visualization demonstrates that the model effectively disentangles behavioral signals and maps them into a representation space where LLM agent behaviors are clearly distinguishable.

\subsection{Identifying LLM Agent Identity}
Beyond distinguishing behavioral patterns, we next investigate whether such traffic fingerprints can also reveal the identity of the underlying LLM agent. As noted earlier, we select the 50 most frequently used \gpts{} from the official store, ensuring diversity in both capabilities and domains while excluding duplicates and low-functionality agents. For each agent, we generate 150 diverse tailored functional prompts and collect interaction traffic through our automated simulation framework, yielding a balanced and representative dataset. The model is then evaluated on agent classification under both closed- and open-world scenarios, with performance compared between using primary-flow and mixed-flow traffic (Figure~\ref{fig:main-4-traffic_result}C).

The experimental results demonstrate the effectiveness of fingerprinting agent identities by user-agent interaction traffic. In the closed-world scenario with primary-flow, the model attains a macro-F1 of 0.8244, with most agents exceeding 0.85. Performance declines only slightly to 0.8118 in the open-world scenario, indicating strong generalization to unseen agents. Incorporating mixed-flow traffic yields further improvements, reaching 0.8660 macro-F1 in the closed-world and 0.8477 in the open-world. This gain is largely attributable to distinctive third-party redirection patterns present in certain agents’ workflows. Figure~\ref{fig:main-4-traffic_result}C illustrates these trends by comparing per-agent F1 distributions across the four settings. Each point represents an individual agent’s score, while violin plots indicate distribution density and dashed lines mark quartiles. In all settings, most agents achieve F1 scores above 0.8, with mixed-flow consistently yielding higher medians and shorter lower tails, reflecting more stable and robust classification. 
While most agents are highly identifiable, classification performance varies with their traffic fingerprint distinctiveness. Agents with unique third-party API integrations or distinctive capability usage (\eg, code execution and multimodal generation) are consistently classified with high accuracy. In contrast, agents exhibiting common traffic patterns (\eg, plain-text, general-purpose agents) are more prone to misclassification, often being confused with one another.

We also visualize the learned representations using t-SNE, as shown in Figure~\ref{fig:main-4-traffic_result}B. At the raw input traffic stage, the embedding space is highly dispersed, with agent identities overlapping and intermingling in a disordered manner. After initial feature extraction, the embeddings show some degree of separation, but remain largely entangled. Unlike behavioral classification tasks, where early-stage clusters typically emerge, the broader diversity of agent interaction patterns makes fingerprint extraction more challenging, and no clear clustering is observed at this point.

As feature extraction progresses, the embeddings become more compact and form distinct clusters, indicating that the model captures the characteristic traffic fingerprints of individual agents. However, certain categories, particularly those near the center of the t-SNE plot, continue to overlap, reflecting residual ambiguity.

Overall, these results demonstrate that the model effectively learns discriminative fingerprints for most agents, though some similarities in behavioral patterns still lead to occasional misclassification.

\subsection{Analyzing Risk of Privacy Leakage}
Behavioral and identity classification of LLM agents can expose users to privacy risks. Once an agent’s identity is inferred, its task orientation, which often reflects user-specific attributes, can serve as a proxy for revealing those attributes. 
To evaluate this risk, we conduct a privacy correlation analysis using a dataset that provides detailed snapshots of available \gpts{} as of October 23, 2024~\cite{sec_gpts_tracker}. From this dataset, we select 12,432 \gpts{} with more than 1,000 recorded conversations and evaluate their potential to reveal private information across multiple user attributes.

Specifically, we design prompts for Qwen-3-plus to determine whether the use of a given agent disproportionately increases the posterior probability of certain user attributes. The evaluation covers a broad range of attributes, including demographic characteristics (such as gender, marital status, and ethnicity), socioeconomic status (such as occupation, education, and financial status), and sensitive domains (such as health, religion, and political orientation). We distinguish three levels of inference: \textit{direct}, when a concrete attribute value can be reliably inferred; \textit{indicative}, when the posterior probability of a broader attribute category increases without identifying a precise label; and \textit{non-indicative}, when no reliable inference can be made. The complete set of prompts is provided at the end of the Supplementary Materials.

\begin{table*}[]
\centering
\caption{
\textbf{Privacy leakage risks posed by \ours{} across different categories of user information for \gpts{} with over 1{,}000 cumulative conversations (N=12{,}432). }
Inference levels are defined as \textit{direct} (access maps to a specific attribute value), \textit{indicative} (access raises the posterior for a broader attribute class without pinpointing a label), and \textit{non-indicative} (no reliable mapping). 
}
\label{tab:privacy_risk}
\begin{tabular}{ll|ccc}
\hline
\multicolumn{2}{c|}{\textbf{User Information}} & \multicolumn{1}{l}{\textbf{Direct}} & \multicolumn{1}{l}{\textbf{Indicative}} & \multicolumn{1}{l}{\textbf{Non-indicative}} \\ \hline
\multicolumn{1}{l|}{\textbf{}} & \textbf{Gender} & 88 & 191 & 12,153 \\
\multicolumn{1}{l|}{\textbf{Demographics}} & \textbf{Marital Status} & 41 & 141 & 12,250 \\
\multicolumn{1}{l|}{\textbf{}} & \textbf{Ethnicity \& Nationality} & 586 & 1,610 & 10,236 \\ \hline
\multicolumn{1}{l|}{\textbf{}} & \textbf{Occupation} & 3,184 & 6,817 & 2,431 \\
\multicolumn{1}{l|}{\textbf{Socioeconomic Status}} & \textbf{Education Background} & 252 & 8,087 & 4,093 \\
\multicolumn{1}{l|}{\textbf{}} & \textbf{Financial Status} & 57 & 1,524 & 10,851 \\ \hline
\multicolumn{1}{l|}{\textbf{}} & \textbf{Health Status} & 110 & 346 & 11,976 \\
\multicolumn{1}{l|}{\textbf{Sensitive Information}} & \textbf{Religious Belief} & 159 & 113 & 12,160 \\
\multicolumn{1}{l|}{\textbf{}} & \textbf{Political Orientation} & 25 & 85 & 12,322 \\ \hline
\end{tabular}
\end{table*}

The results, summarized in Table~\ref{tab:privacy_risk}, reveal a heterogeneous landscape of privacy risks. Categories with strong correlations, such as occupation and educational background, exhibit broad vulnerability. This outcome reflects common deployment practices of LLM agents: many are designed to support work-related activities, and their integration into professional workflows naturally aligns with task execution in occupational contexts. Conversations in these settings often involve domain-specific terminology and problem-solving routines, which provide strong signals of users’ occupational roles or educational training. The close connection between occupation and education further amplifies their visibility in conversational traces.

Although relatively few agents disclose attributes such as political orientation, religious belief, or certain health conditions, the associated risks remain substantial. Adversaries could exploit highly correlated agents to construct targeted monitoring sets, enabling them to track or discriminate against users with these attributes. Crucially, every category of user information is linked to at least some agents that leak sensitive attributes, indicating that privacy risks are not confined to isolated cases but are pervasive across the landscape. Overall, these findings highlight that all dimensions of user information carry tangible privacy risks, whether through broad inference or targeted exploitation.

To ground our analysis in a concrete example, we focus on occupation as a representative case study. Occupation is a central component of socioeconomic status~\cite{privacy_occ1, privacy_occ2, FTC2024}, supported by comprehensive datasets and well-established taxonomies~\cite{onet2023}. These resources facilitate rigorous evaluation~\cite{llm_occ2023, llm_occ2024, llm_occ_ms2024} and make occupational profiling both analytically tractable and societally significant.

\subsection{Occupational Profiling Experiments}

We evaluate the proposed zero-shot occupation profiling method on both real-world and synthetic users to quantify the risk of inferring sensitive attributes from cross-agent usage patterns. 
Both experiments employ the O*NET 27.2 database~\cite{onet2023} as the occupational taxonomy. O*NET is a comprehensive classification system maintained by the U.S. Department of Labor, Employment and Training Administration (USDOL/ETA), defining 923 detailed occupations with standardized titles, descriptions, tasks, and Detailed Work Activities (DWAs). Tasks specify concrete job duties, whereas DWAs represent broader functional activities that may span multiple occupations. This taxonomy provides a consistent framework for mapping the functional capabilities of LLM-based agents to work activities associated with different professions.

Following~\cite{GPT_labor_science}, we use the concept of \emph{exposure}, defined as the proportion of an occupation's DWAs that can be significantly assisted or automated by LLMs, to quantify the potential impact of LLMs on each occupation. The same work also constructs an occupation network based on DWA similarity and applied the Louvain community detection algorithm, yielding 11 distinct occupation categories: \textit{Managers}, \textit{Clerks and Services}, \textit{Technologists}, \textit{Architects and Engineers}, \textit{Scientists and Researchers}, \textit{Medical Workers}, \textit{Legal Services}, \textit{Teachers}, \textit{Arts, Media, and Entertainment}, \textit{Operators}, and \textit{Machinists}.

As defined earlier, the goal of the adversary is to infer a victim's occupation category from the set of identified agents with which the victim interacts. To evaluate this threat, we generate virtual user profiles using the generative capabilities of LLMs and conduct a real-world user study to capture actual agent usage preferences.

For virtual user generation, we iterate over all 923 occupations in O*NET. For each occupation, we provide its title and a brief description from O*NET to DeepSeek-V3, prompting it to simulate the daily workflow of a professional in that role. Importantly, we exclude O*NET tasks and DWAs, which are reserved for our inference method, ensuring that the generated usage patterns rely only on the model’s general knowledge of the occupation. The simulated workday, together with a list of closed-world agents, is then presented to DeepSeek-R1, which selects and ranks the agents most likely to be used during that workday. This step leverages the model’s reasoning capacity to produce more accurate rankings. The outcome is a ranked agent list for each occupation, forming the basis of the virtual user profiles. To introduce variation, we repeat the process with different random seeds, generating multiple virtual users per occupation.

\begin{table*}[]
\centering
\caption{
    \textbf{Performance of occupation inference for high-exposure virtual users (exposure $>$ 0.4).} 
    Results are computed on 3{,}306 virtual users generated from 551 O*NET occupations whose LLM exposure exceeds 0.4. 
    Agent usage profiles are randomly generated to reflect the daily workflows of each occupation. 
    ACC@$N$ denotes the proportion of samples where the correct label is among the top $N$ predictions. 
    FNR is the proportion of misclassified users. 
    Our approach achieves a Top-3 accuracy of 73.9\%, indicating a substantial profiling risk for high-exposure occupations.
}
\label{tab:infer_virtual_user}
\begin{tabular}{lrrrrr}
\hline
\textbf{Occupation Category} &
  \multicolumn{1}{c}{\textbf{Users}} &
  \multicolumn{1}{c}{\textbf{ACC@1}} &
  \multicolumn{1}{c}{\textbf{ACC@2}} &
  \multicolumn{1}{c}{\textbf{ACC@3}} &
  \multicolumn{1}{c}{\textbf{FNR}} \\ \hline
\textbf{Managers}                       & 21.23\% & 80.34\% & 89.74\% & 93.59\% & 6.41\%  \\
\textbf{Clerks and Services}            & 17.97\% & 13.30\% & 60.77\% & 69.02\% & 30.98\% \\
\textbf{Technologists}                  & 6.72\%  & 45.05\% & 66.22\% & 77.03\% & 22.97\% \\
\textbf{Architects and Engineers}       & 9.98\%  & 40.61\% & 83.33\% & 93.03\% & 6.97\%  \\
\textbf{Scientists and Researchers}     & 11.98\% & 3.28\%  & 36.87\% & 73.74\% & 26.26\% \\
\textbf{Medical Workers}                & 12.52\% & 82.85\% & 87.68\% & 91.06\% & 8.94\%  \\
\textbf{Legal Services}                 & 1.27\%  & 0.00\%  & 2.38\%  & 28.57\% & 71.43\% \\
\textbf{Teachers}                       & 10.71\% & 20.62\% & 26.55\% & 31.64\% & 68.36\% \\
\textbf{Arts, Media, and Entertainment} & 5.63\%  & 30.11\% & 41.40\% & 52.15\% & 47.85\% \\
\textbf{Operators}                      & 0.73\%  & 0.00\%  & 0.00\%  & 8.33\%  & 91.67\% \\
\textbf{Machinists}                     & 1.27\%  & 0.00\%  & 7.14\%  & 14.29\% & 85.71\% \\ \hline
\textbf{Total}                          & 13.69\% & 41.20\% & 63.43\% & 73.90\% & 26.10\% \\ \hline
\end{tabular}%
\end{table*}

We first study the relationship between LLM exposure and profiling accuracy, using the average of human- and GPT-4-labeled exposure scores reported in~\cite{GPT_labor_science}. As shown in Figure~\ref{fig:main-5-occupation_result}B, accuracy increases steadily for occupations with exposure below 0.4 and plateaus at around 0.7 for those above 0.4. This trend supports the hypothesis that occupations more exposed to LLMs, and thus more likely to rely on LLM-based agents, leak greater amounts of information through usage behaviors and are consequently more vulnerable to profiling. On this basis, we designate occupations with exposure above 0.4 as high-exposure.

Table~\ref{tab:infer_virtual_user} reports profiling performance for 3{,}306 virtual users derived from 551 high-exposure occupations. The method achieves a top-3 accuracy of 73.9\%, indicating substantial leakage risk for professions with greater LLM exposure. Performance varies markedly across categories. \textit{Managers} and \textit{Medical Workers} achieve over 80\% top-1 accuracy, reflecting highly distinctive usage patterns, whereas \textit{Legal Services}, \textit{Operators}, and \textit{Machinists} show near-zero top-1 accuracy, due to weaker alignment between their functional requirements and the available closed-world agents. Intermediate performance appears in \textit{Clerks and Services} and \textit{Technologists}, where higher-order predictions (top-2 or top-3) recover many top-1 misclassifications. Extending the evaluation to all 923 occupations reduces top-3 accuracy to 58.9\%, primarily due to the large number of \textit{Operator} and \textit{Machinist} roles with accuracies below 10\%. These results underscore that high-exposure occupations are inherently easier to profile, due to their stronger functional alignment with specific LLM agents and more distinctive cross-agent usage signatures.

\begin{table*}[]
\centering
\caption{
    \textbf{Performance of occupation inference for real-world users from LLM agent usage patterns.} 
    Based on a survey of 49 participants selecting and ranking their top 5 most frequently used GPTs. 
    ACC@$N$ and FNR follow the same definitions as in Table~\ref{tab:infer_virtual_user}. 
    The method achieves a Top-3 accuracy of 69.1\%.
    The accuracies may take non-binary values because each participant's responses are resampled multiple times by the error rate during the agent identification stage. 
}
\label{tab:infer_real_user}
\begin{tabular}{lrrrrr}
\hline
\textbf{Occupation Category} &
  \multicolumn{1}{c}{\textbf{Users}} &
  \multicolumn{1}{c}{\textbf{ACC@1}} &
  \multicolumn{1}{c}{\textbf{ACC@2}} &
  \multicolumn{1}{c}{\textbf{ACC@3}} &
  \multicolumn{1}{c}{\textbf{FNR}} \\ \hline
\textbf{Managers}                       & 13 & 58.74\% & 69.83\% & 77.42\% & 22.58\% \\
\textbf{Clerks and Services}            & 8  & 15.95\% & 48.79\% & 59.34\% & 40.66\% \\
\textbf{Technologists}                  & 12 & 53.46\% & 63.44\% & 70.84\% & 29.16\% \\
\textbf{Scientists and Researchers}     & 1  & 0.30\%  & 78.10\% & 87.80\% & 12.20\% \\
\textbf{Medical Workers}                & 6  & 90.30\% & 97.25\% & 98.43\% & 1.57\%  \\
\textbf{Legal Services}                 & 1  & 0.00\%  & 6.90\%  & 77.50\% & 22.50\% \\
\textbf{Teachers}                       & 1  & 92.00\% & 96.00\% & 97.70\% & 2.30\%  \\
\textbf{Arts, Media, and Entertainment} & 4  & 25.35\% & 29.25\% & 30.95\% & 69.05\% \\
\textbf{Operators}                      & 2  & 0.00\%  & 0.00\%  & 35.70\% & 64.30\% \\
\textbf{Machinists}                     & 1  & 0.20\%  & 2.90\%  & 5.80\%  & 94.20\% \\ \hline
\textbf{Total}                          & 49 & 46.29\% & 60.08\% & 69.10\% & 30.90\% \\ \hline
\end{tabular}%
\end{table*}

For the real-world user study, we recruit 49 participants with prior LLM usage experience through the Prolific~\footnote{\url{https://www.prolific.com}} platform. Each participant is presented with 55 LLM agents with distinct functionalities and asked to select those they would likely use in daily work, rank their top five, and indicate relative usage frequencies. As reported in Table~\ref{tab:infer_real_user}, our method achieves a top-3 accuracy of 69.1\%, closely matching the 73.9\% observed for high-exposure virtual users in Table~\ref{tab:infer_virtual_user}. This consistency suggests that both settings capture comparable leakage patterns, thereby mutually reinforcing the validity of the results.

In both datasets, \textit{Managers} and \textit{Medical Workers} yield high accuracy, reflecting distinctive and well-aligned usage patterns, whereas \textit{Legal Services}, \textit{Operators}, and \textit{Machinists} remain difficult to profile. Notable differences include higher accuracy for \textit{Medical Workers} in the real-world study, likely because participants’ workflows align closely with the available agents, and lower performance for some mid-range categories (\eg, \textit{Clerks and Services}), possibly due to greater variability in real-world usage habits and smaller sample sizes.

After comparing overall performance between high-exposure virtual and real users, we further analyze how profiling accuracy varies with the amount of observed cross-agent usage. Figure~\ref{fig:main-5-occupation_result}C shows that accuracy increases consistently as more agents become visible in both settings, across all top-$K$ metrics. This trend indicates that access to a broader range of agent-usage information enables the attacker to construct a more complete behavioral profile, thereby improving occupation inference performance.

After analyzing the effect of agent visibility, we next examine how the accuracy of the preceding agent identification stage shapes downstream occupation inference. As shown in Figure~\ref{fig:main-5-occupation_result}D, profiling accuracy increases almost linearly with identification accuracy across all top-$K$ metrics, demonstrating that reliable recognition of a user’s agent usage is a prerequisite for effective occupation inference. In subsequent experiments, we fix identification accuracy at 0.866, corresponding to the empirical performance obtained in the agent identity classification experiment. This level of precision ensures that the attacker possesses sufficiently reliable agent-usage information, thereby making large-scale and accurate occupation profiling feasible.

\section{Prompts used in the Experiments}
During the experiments, we mainly use Qwen-3-plus to annotate the privacy leakage risk for LLM agents and the exposure of DWAs to these agents, \ie, the alignment between the functionality of LLM agents and DWAs. 
We employ Deepseek-R1 and Deepseek-V3 for generating virtual user profiles based on LLM agent usage. The prompts used in these processes are outlined in this section.

\subsection{The Prompt for DWA-LLM Agent Exposure Annotation}
\begin{tcolorbox}[title=Annotation of DWA Exposure to LLM Agents, breakable]
\textbf{Task}
An LLM agent is a system built on a base Large Language Model (LLM), enhanced with specific instructions, tools, or workflows to perform particular types of tasks. LLM agents may be equipped with: Domain-specific knowledge or fine-tuning, Access to third-party APIs, Web browsing for real-time information, File analysis capabilities, Code execution, and Image generation.
You are attempting to complete a specific Detailed Work Activity (DWA)-a clearly defined, job-relevant activity that represents a meaningful unit of work commonly performed in one or more occupations. You are given access to a specific LLM agent. This agent may or may not be explicitly designed for your DWA.
Your goal is to determine whether the LLM agent would be helpful or helpless in completing the activity, based on its name, description, and example use cases.

\textbf{Rubric}
Use the following criteria to make your decision:
\begin{itemize}\setlength{\itemsep}{0 pt}\setlength{\parsep}{0 pt}%
\item Helpful (1): 
    The LLM agent includes relevant capabilities or domain-specific knowledge that align with the activity. It can provide direct or partial assistance in completing the activity. Example: A ``Resume Analyzer'' agent would be helpful for a Human Resources Manager evaluating candidate CVs.
\item Helpless (0): 
    The LLM agent is clearly unrelated to the activity and offers no meaningful advantage over a base LLM. This includes cases where:
    The agent is intended for a different domain or activity type.
    The activity requires physical or real-world actions that a virtual assistant cannot perform or meaningfully assist with (\eg, washing dishes, driving a vehicle).
    The activity lies outside the cognitive or functional scope of an LLM agent, even with tools (\eg, representing someone in court, performing a surgical procedure).
    Example: A ``Math Teaching Assistant'' agent would be helpless for summarizing a legal contract.
\end{itemize}

\textbf{Output Format}
Return a single binary score with no explanation or additional text:
\begin{itemize}\setlength{\itemsep}{0 pt}\setlength{\parsep}{0 pt}%
\item 0: The LLM agent is helpless for completing the activity.
\item 1: The LLM agent is helpful for completing the activity.
\end{itemize}

\textbf{Input Fields}
\begin{itemize}\setlength{\itemsep}{0 pt}\setlength{\parsep}{0 pt}%
  \item Agent Name: \texttt{<\$name>}
  \item Description: \texttt{<\$description>}
  \item Example Prompts: \texttt{<\$example\_prompts>}
  \item Detailed Work Activity (DWA): \texttt{<\$DWA>}
  \item Occupations and corresponding tasks related to the DWA: \texttt{<\$occupation, \$task>}
\end{itemize}

\end{tcolorbox}

\subsection{The Prompt for Privacy Leakage Risk Analysis}
\begin{tcolorbox}[title=Privacy Leakage Risk Analysis, breakable]
\textbf{Role}
You are an AI security expert. Determine whether the observable act of accessing a specific LLM Agent, without any conversation content, could reveal private user attributes.

\textbf{Access-only Evidence Rule}
\begin{itemize}\setlength{\itemsep}{0 pt}\setlength{\parsep}{0 pt}%
  \item The only evidential signal is the fact of access to the Agent.
  \item Treat titles and descriptions as indicators of the Agent's scope and specialization.
  \item Do not use example prompts as evidence for inference. They are illustrative and may reflect edge cases or hypothetical users.
  \item If an example prompt mentions a narrow attribute, but the Agent itself is general-purpose, do not infer that attribute from access alone.
\end{itemize}

\textbf{Core Definition}
Privacy inference risk exists if observing only that a user accessed the Agent yields a statistically meaningful posterior lift for a private attribute relative to its prior. Assess whether P(attribute \textbar access) is materially greater than P(attribute). Ignore conversation content and any external data.

\textbf{Standardized Privacy Categories to Evaluate}
\begin{itemize}\setlength{\itemsep}{0 pt}\setlength{\parsep}{0 pt}%
  \item Gender
  \item Marital Status
  \item Ethnicity, Nationality, or Linguistic Background
  \item Religious Belief
  \item Political Orientation
  \item Occupation
  \item Education Background
  \item Financial Status
  \item Health Status
\end{itemize}

\textbf{Scoring Scheme}
Use three mutually exclusive levels, from strongest to weakest signal.
\begin{itemize}\setlength{\itemsep}{0 pt}\setlength{\parsep}{0 pt}%
  \item \textbf{inference\_level = direct}: Access maps to a specific attribute value or a very narrow set of values (large posterior lift).\\
  \emph{Examples:} bitcoin advisor $\Rightarrow$ crypto asset involvement; Quran agent $\Rightarrow$ Islam; Japanese script agent $\Rightarrow$ Japanese language use; medical literature retriever $\Rightarrow$ research roles; breast-cancer-specific treatment agent $\Rightarrow$ breast oncology condition class.
  \item \textbf{inference\_level = indicative}: Access is domain-related and raises the posterior for a broader attribute class, but does not pinpoint a specific label.\\
  \emph{Examples:} general doctor consultation agent $\Rightarrow$ illness (unspecified); document research helper $\Rightarrow$ paper-reading roles; general DNA variant interpreter $\Rightarrow$ medical-genetics interest (no gender/disease implied).
  \item \textbf{inference\_level = non-indicative}: No reliable mapping from access to the attribute; posterior lift is negligible or unsupported.
\end{itemize}

\textbf{Output Format}
Return a single JSON object. Do not include explanations outside JSON. 
\begin{quote}\small\ttfamily
[\\
\hspace*{2em}\{ "category": "Gender", "is\_related": true, "reason": "xxx", "inference\_level": "indicative" \},\\
\hspace*{2em}\ldots\\
]
\end{quote}

\textbf{Additional Rules}
\begin{itemize}\setlength{\itemsep}{0 pt}\setlength{\parsep}{0 pt}%
  \item \texttt{is\_related} is \texttt{true} when any meaningful posterior lift exists; otherwise \texttt{false}.
  \item \texttt{reason} is 1-2 sentences grounded in the Agent's function and typical user base. Do not rely on conversation content or example prompts.
  \item Use exactly the category labels provided; do not add, rename, split, or nest categories.
  \item Output only categories where \texttt{is\_related = true}; optionally include up to two \texttt{non-indicative} categories to clarify misconceptions.
  \item If multiple levels could apply, choose the strongest level justified by access-only evidence.
\end{itemize}

\textbf{Input Fields}
\begin{itemize}\setlength{\itemsep}{0 pt}\setlength{\parsep}{0 pt}%
  \item Agent Name: \texttt{<\$name>}
  \item Description: \texttt{<\$description>}
  \item Example Prompt: \texttt{<\$example\_str>}
\end{itemize}
\end{tcolorbox}

\subsection{The Prompt for LLM Agents Usage Profile Generation}
Virtual users are generated through a two-step process: (1) occupational workflow simulation, which models a typical workday scenario for a given occupation, and (2) agent selection, in which five relevant agents are selected based on the simulated workflow.

\begin{tcolorbox}[title=LLM Agents Usage Profile Generation, breakable]
\textbf{Step 1: Occupational Workflow Simulation}\\
You are an AI language model tasked with role-playing as a professional in a specific occupation. Your goal is to simulate a typical workday in that role and identify which LLM agents would best assist you in completing your tasks.

An LLM agent is a system built on top of a base Large Language Model, augmented with specialized instructions, tools, or workflows tailored for particular types of tasks. These agents may be equipped with: Domain-specific knowledge or fine-tuning, Access to third-party APIs, Web browsing for real-time information, File analysis capabilities, Code execution, Image generation.

\textbf{Occupation Information}\\
Title: \texttt{<\$occ\_title>}\\
Description: \texttt{<\$occ\_desc>}  

\textbf{Instruction}
Your task is to imagine one specific workday---from beginning to end. Describe what you worked on that day, the kinds of tasks you performed, what problems you encountered, and what goals you tried to achieve. Please write one paragraph narrating this imagined day of work, focusing on the practical activities and decisions you engaged in.
In later steps (not yet), you will select suitable LLM agents to assist with specific parts of your work.
\\
\\
\textbf{Step 2: Agent Selection}\\
Based on the workday you just described, select the 5 most helpful LLM agents that would best support you in completing your tasks, sorted by the frequency you will use the LLM agent.
Consider which agents would be useful for solving specific problems, speeding up your workflow, or improving the quality of your output. Please return your selection as a list of 5 agent names, each followed by a short explanation (1-2 sentences) of why it would be helpful for your work. Here is the list of available agents with their name and a brief description: Available agents: \texttt{<\$agent\_list\_block>}
\end{tcolorbox}

\end{document}